\documentclass[preprint,prb]{revtex4-1}

\usepackage{amsmath,amsfonts,amssymb}
\usepackage{graphicx,pstricks,caption}
\usepackage[position=top]{subfig}
\usepackage{graphbox}
\usepackage{float}

\usepackage{epstopdf}

\begin{document}

\title{Chemical functionalization, electronic and dielectric properties of hybrid organic-tin layers}
\author{Andre Oliveira}
\affiliation{Universidade Federal de Goi\'as, Institute of Physics, Campus Samambaia, 74690-900 Goi\^ania, Goi\'as, Brazil}
\author{Andreia Luisa da Rosa}
\affiliation{Universidade Federal de Goi\'as, Institute of Physics, Campus Samambaia, 74690-900 Goi\^ania, Goi\'as, Brazil}
\affiliation{Bremen Center for Computational Materials Science, University of Bremen, Am Fallturm 1, 28359 Bremen, Germany}
\author{Thomas Frauenheim}
\affiliation{Bremen Center for Computational Materials Science, University of Bremen, Am Fallturm 1, 28359 Bremen, Germany}
\affiliation{Computational Science Research Center, No.10 East Xibeiwang Road, Beijing 100193 and Computational Science and Applied Research Institute Shenzhen, China.}

\begin{abstract}

  Band gap tuning and dielectric properties of small organic ligands
  adsorbed on tin monolayers (stanene) have been
  investigated using first-principles calculations. Charge density
  analysis using density-functional theory shows that the ligands are chemisorbed on stanene 
  and some of the groups can open a band gap in the originally
  metallic statene. Furthermore many-body GW calculations demonstrate that the dielectric  properties of bare and ligand adsorbed stanene have a large
  anisotropy. Our findings of a finite gap opens a path for
  rational  theoretical design of functionalized two-dimensional stanene.

\end{abstract}


\maketitle

\section{Introduction}

 Graphene in its honeycomb structure has a zero gap, with conducion
 and valence bands being degenerate at K and K' points, forming Dirac
 cones with charge carriers which are massless
 fermions\,\cite{Novoselov:04,Geim:07}. Graphene has a metallic
 behavior and open a band gap in graphene has been challenging. In
 general, for the creation of optical and electronic devices
 semiconductor properties are desirable.  Two-dimensional
 nanostructures built from group 14 elements, such as tin, silicon and
 germanium have been the topic of several
 investigations\cite{tang2014stable}. In fact, stanene has already been demonstrated on substrates such as
 Bi$_2$Te$_3$\,\cite{ni2017germanene} and Ag(111)\cite{Nat2019}. From
 the theoretical side, first-principles molecular dynamics have shown
 that bare stanene has a phonon dispersion without imaginary
 frequencies and the monolayer is thermally stable up to 700
 K\cite{peng2016low}.

 Similarly to graphene, stanene has a metallic
 behavior. Some ways of opening a band gap in stanene
 include doping by impurities, defect engineering, functionalization
 with organic and/or biological molecules\,\cite{Zhang_2015,Zhang2016}
 and applied electric field and applied strain\cite{Broek_2014}. A
 promising route for surface functionalization of two-dimensional
 materials was demonstrated using adsorption of organic molecules or
 functional groups, as it has been reported for
 germanene\,\cite{ProgressReports}.  However, experimental results on
 functionalized stanene are still missing.

In this work we have investigated the
electronic and dielectric properties of stanene with small organic
groups using first-principles calculations. We show that the some of
the adsorbed stanene are suitable for opening a band gap in the
metallic stanene.  We also show that the dielectric properties of this
material have a large anisotropy, with absorption in the visible
region, which could render possible applications in optoelectronic
devices.

\section{Methodology}

We employ density functional theory\,\cite{Hohenberg:64,Kohn:65}
within the generalized gradient approximation\,\cite{Perdew:96} and
the projected augmented wave method
(PAW)\,\cite{Bloechel:94,Kresse:99} as implemented in the VASP
code\,\cite{Kresse:99} to investigate the electronic structure of
hybrid statenen monolayers. ($1\times$1) and ($2\times$2) supercells
were employed to simulate full and partial ligand coverages. Forces on
atoms were converged until 10$^{-4}$\,eV/{\AA}. A vacuum region of 10
{\AA} is sufficient to avoid spurious interactions between stanene
layers in neighbouring cells. A (8$\times$8$\times$1) {\bf k}-point
sampling has been used in the calculations of all investigated systems
with an energy cutoff of 400 eV. The calculation of the dielectric
function was performed using the GW method\,\cite{Shishkin:07} with a
(6$\times$6$\times$1) {\bf k}-points mesh.

\section{Results and Discussions}

\subsection{Structural properties}

\begin{figure}[htp!]
  \centering
  \includegraphics[width=6cm,clip = true]{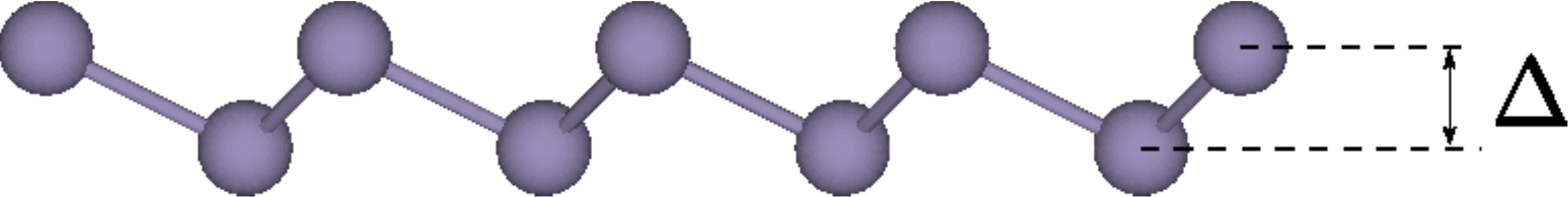}
  \includegraphics[width=6cm, scale=1, clip=true, keepaspectratio]{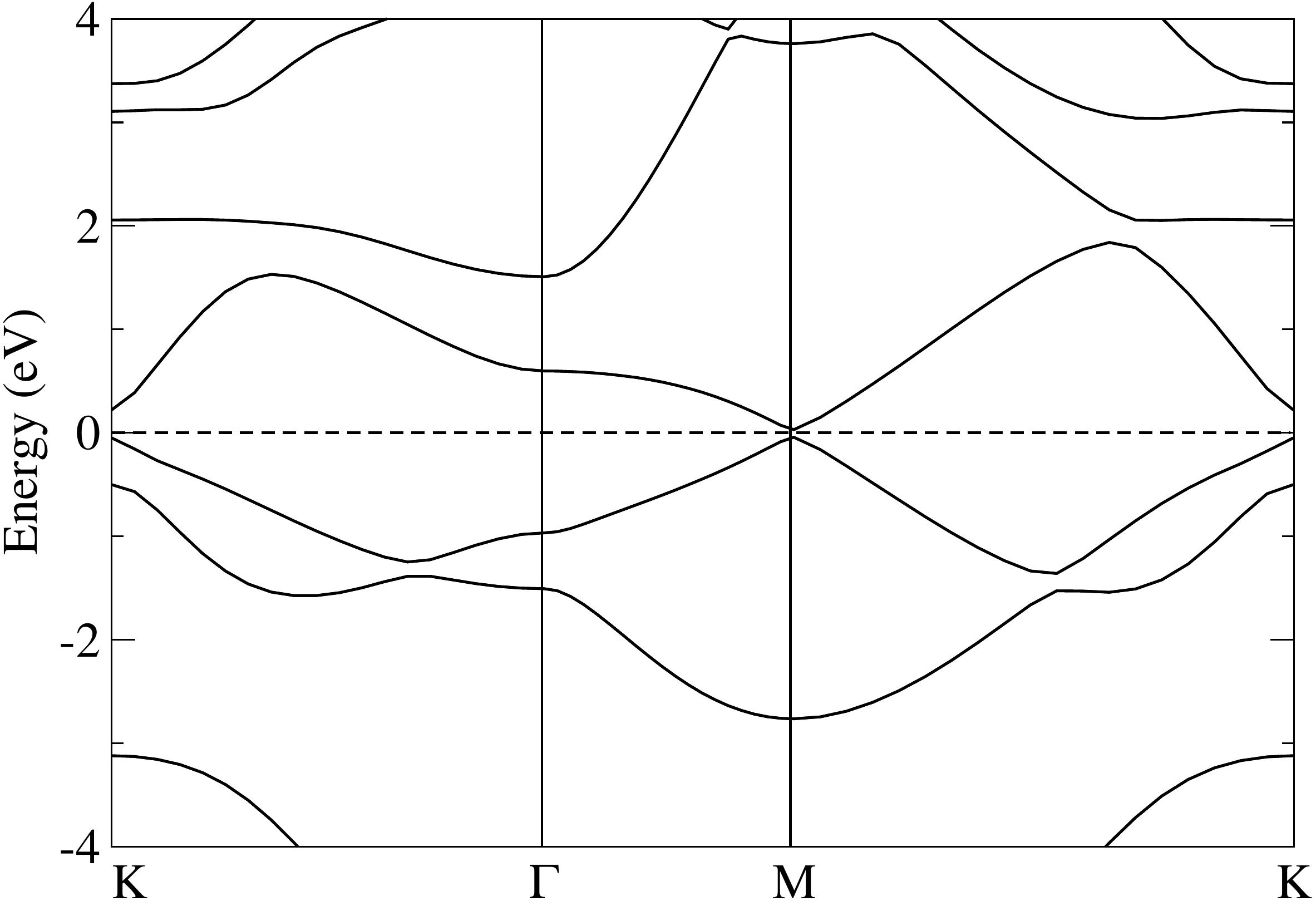}\\
\includegraphics[width=6cm, scale=1, clip = true]{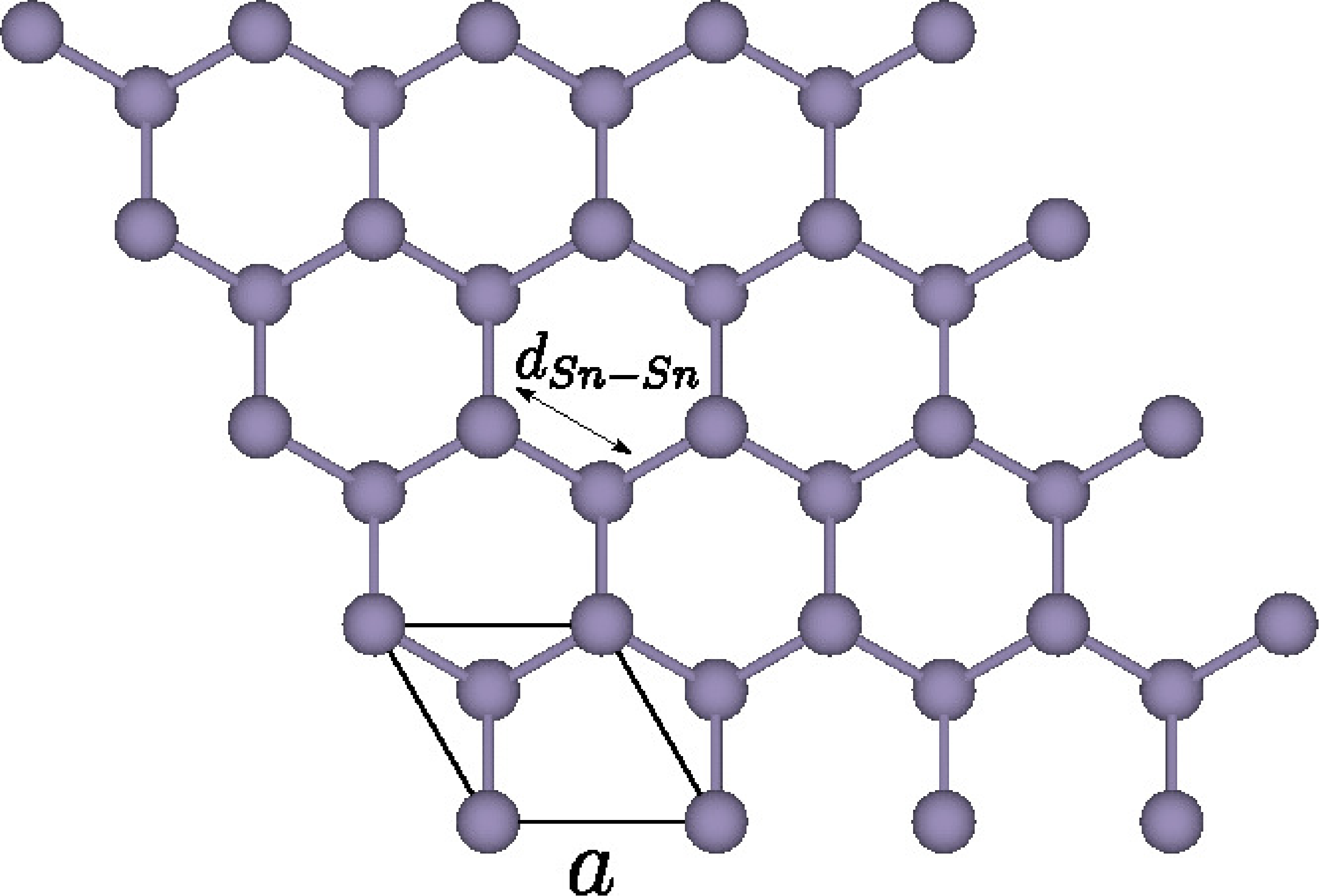}
\includegraphics[width=6cm, scale=1, clip = true, keepaspectratio]{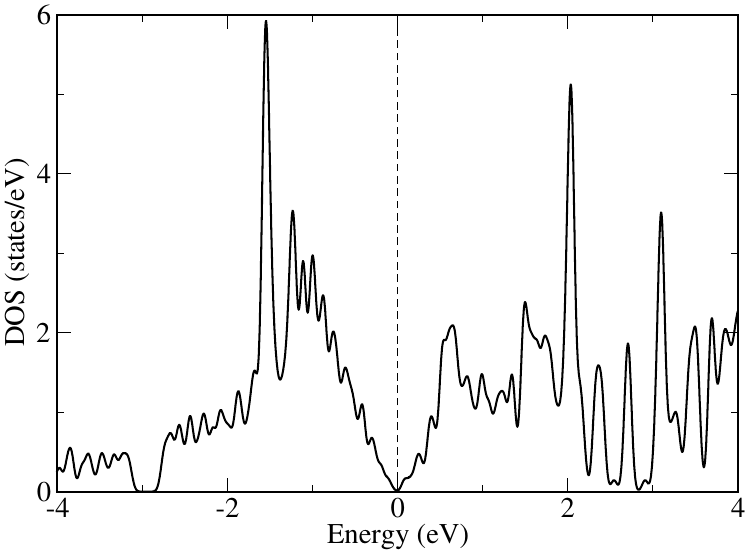}
  \caption{\label{fig:structuresbare} a) top view, b) side view, c) density of states and d) band structure of optimized tin layers within GGA.}
\end{figure}

\begin{figure}[htp!]
  \centering
  \includegraphics[width=4cm]{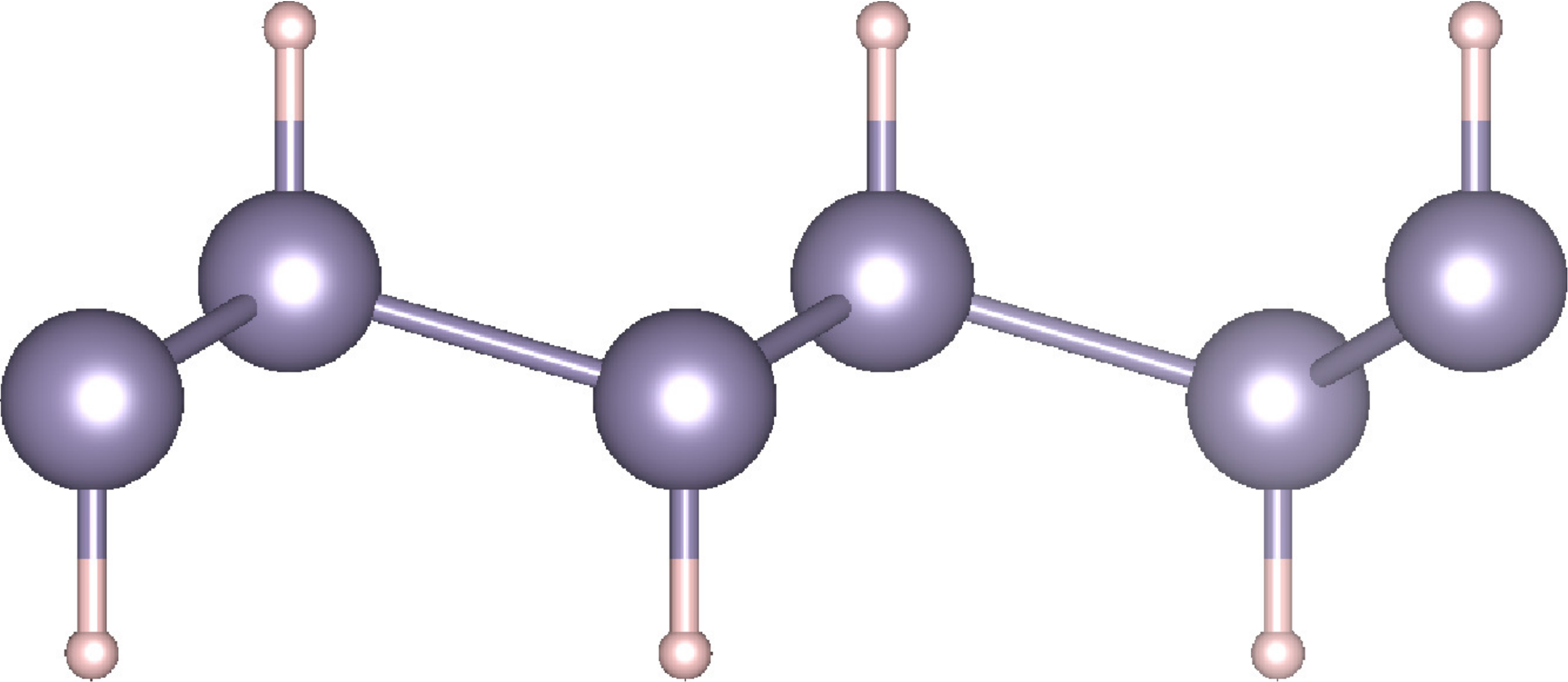}\hfill
  \includegraphics[width=4cm]{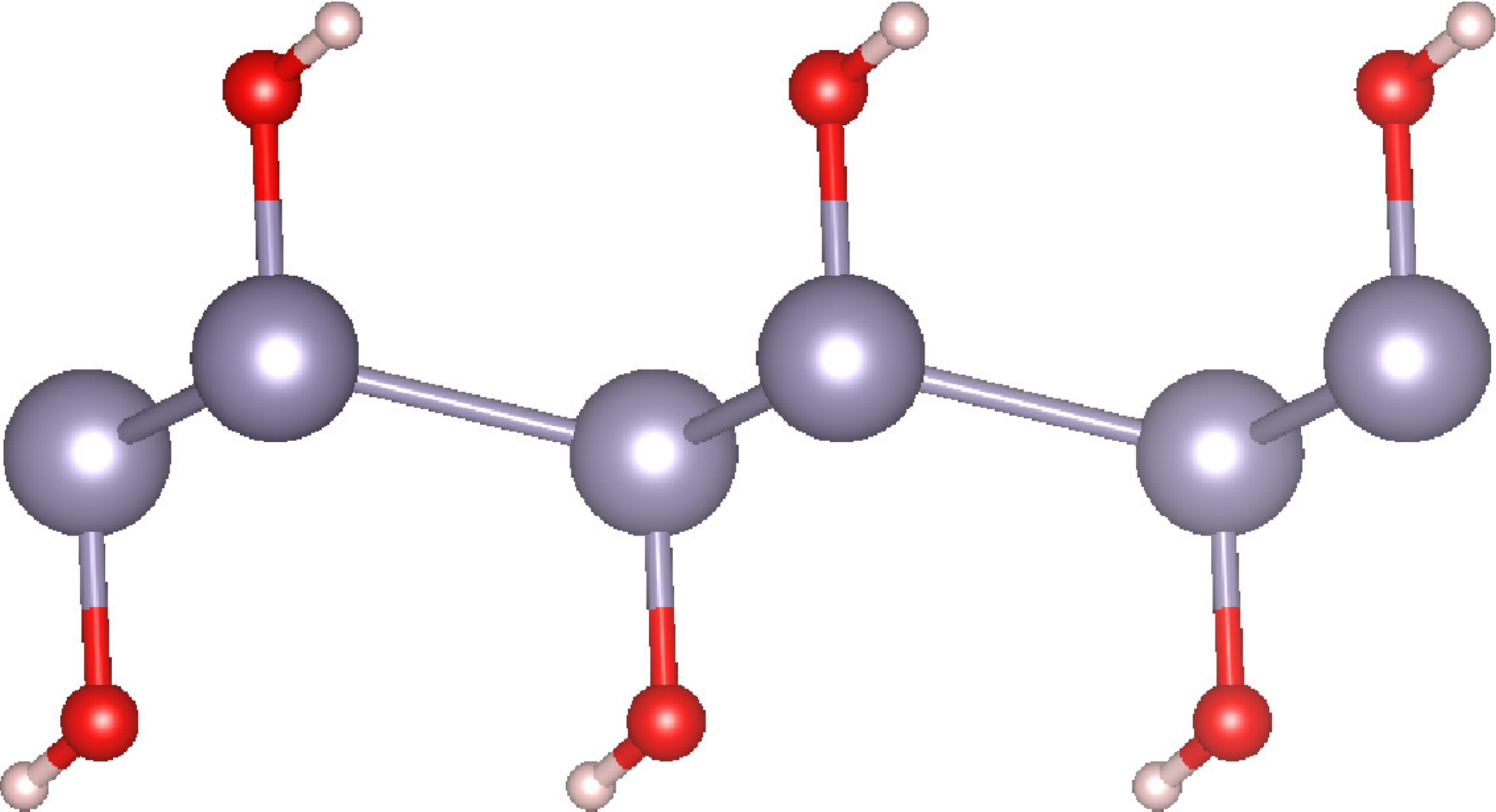}\hfill
  \includegraphics[width=4cm]{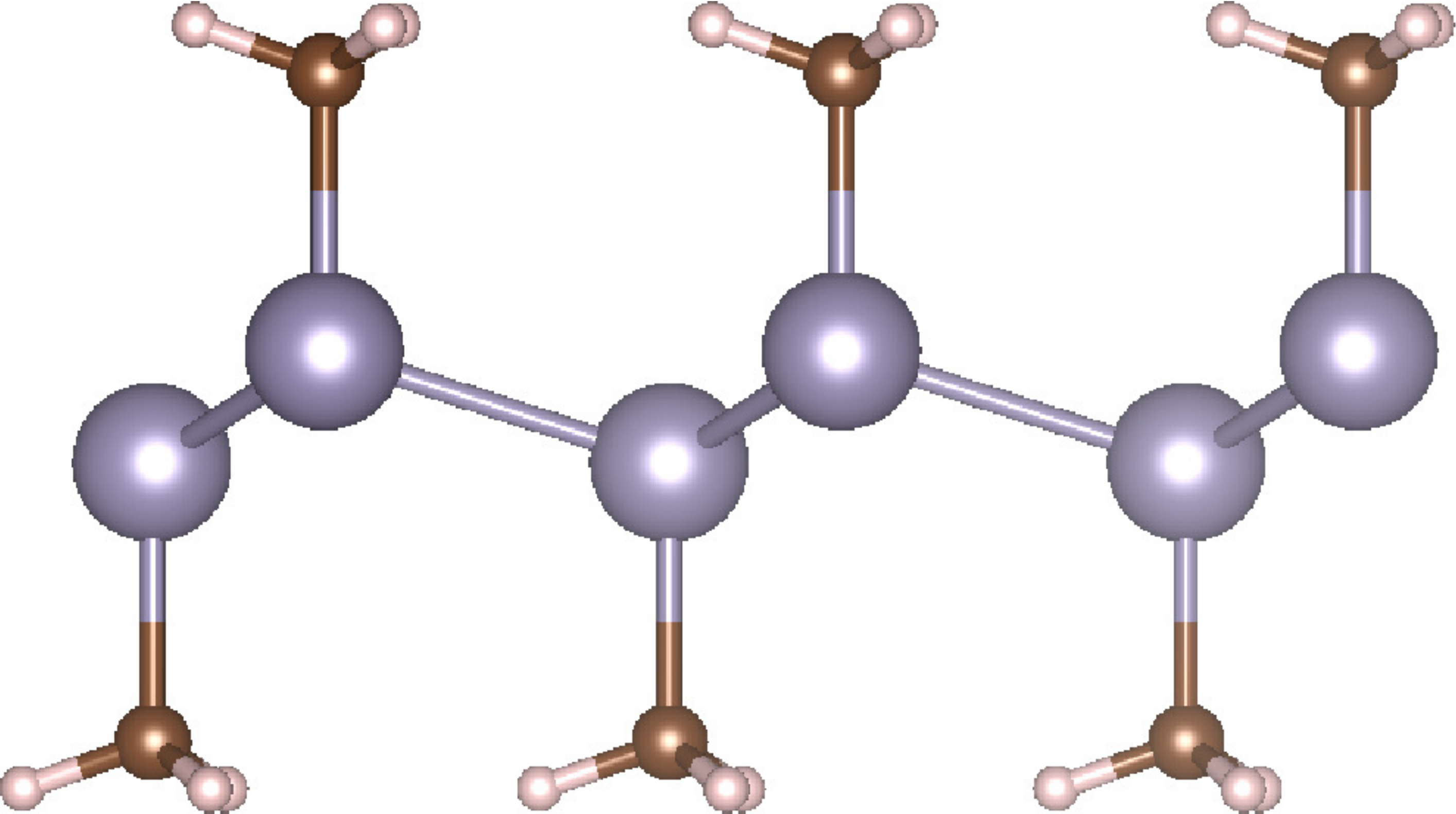}\\
  \includegraphics[width=4cm]{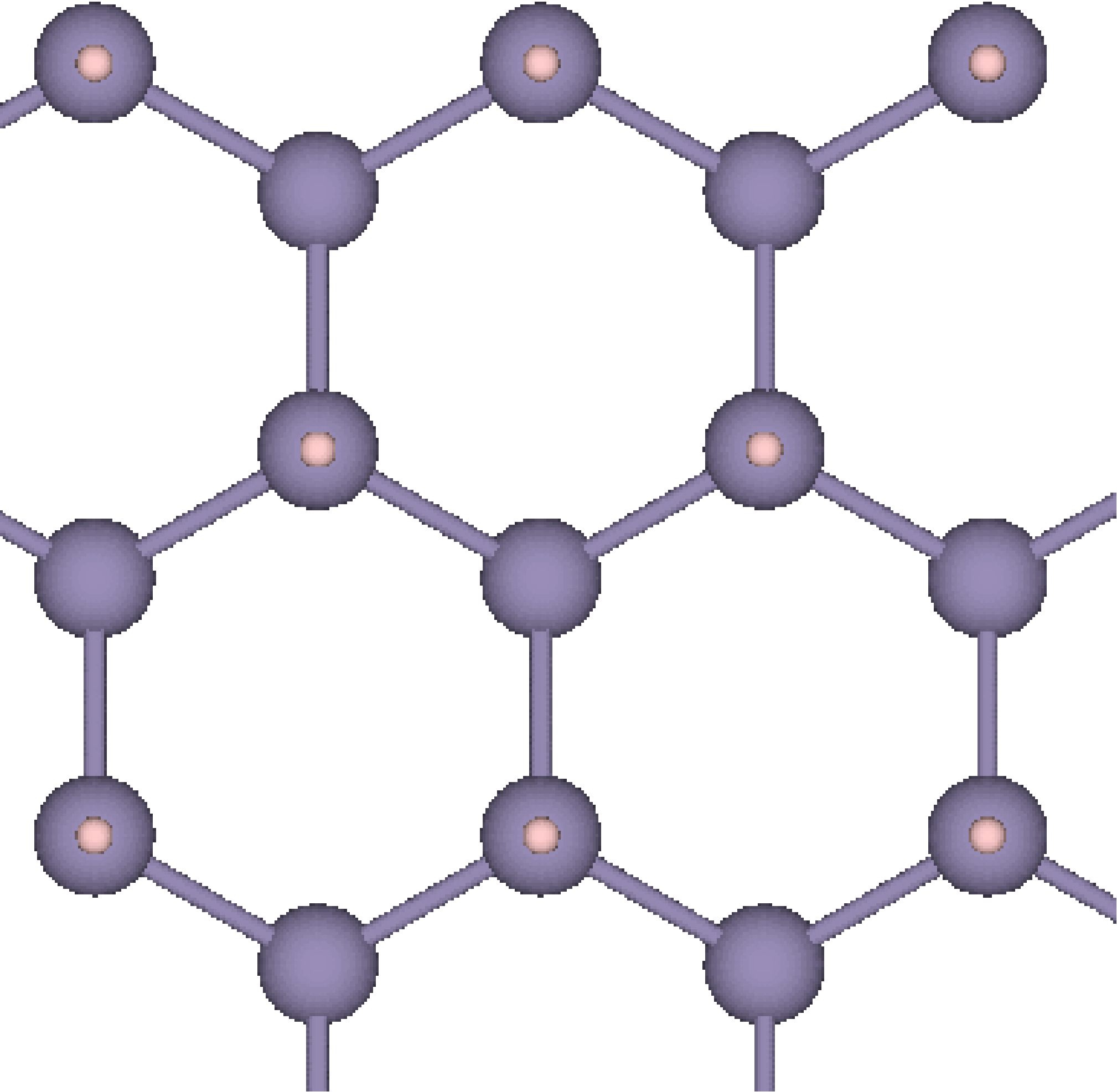}\hfill
  \includegraphics[width=4cm]{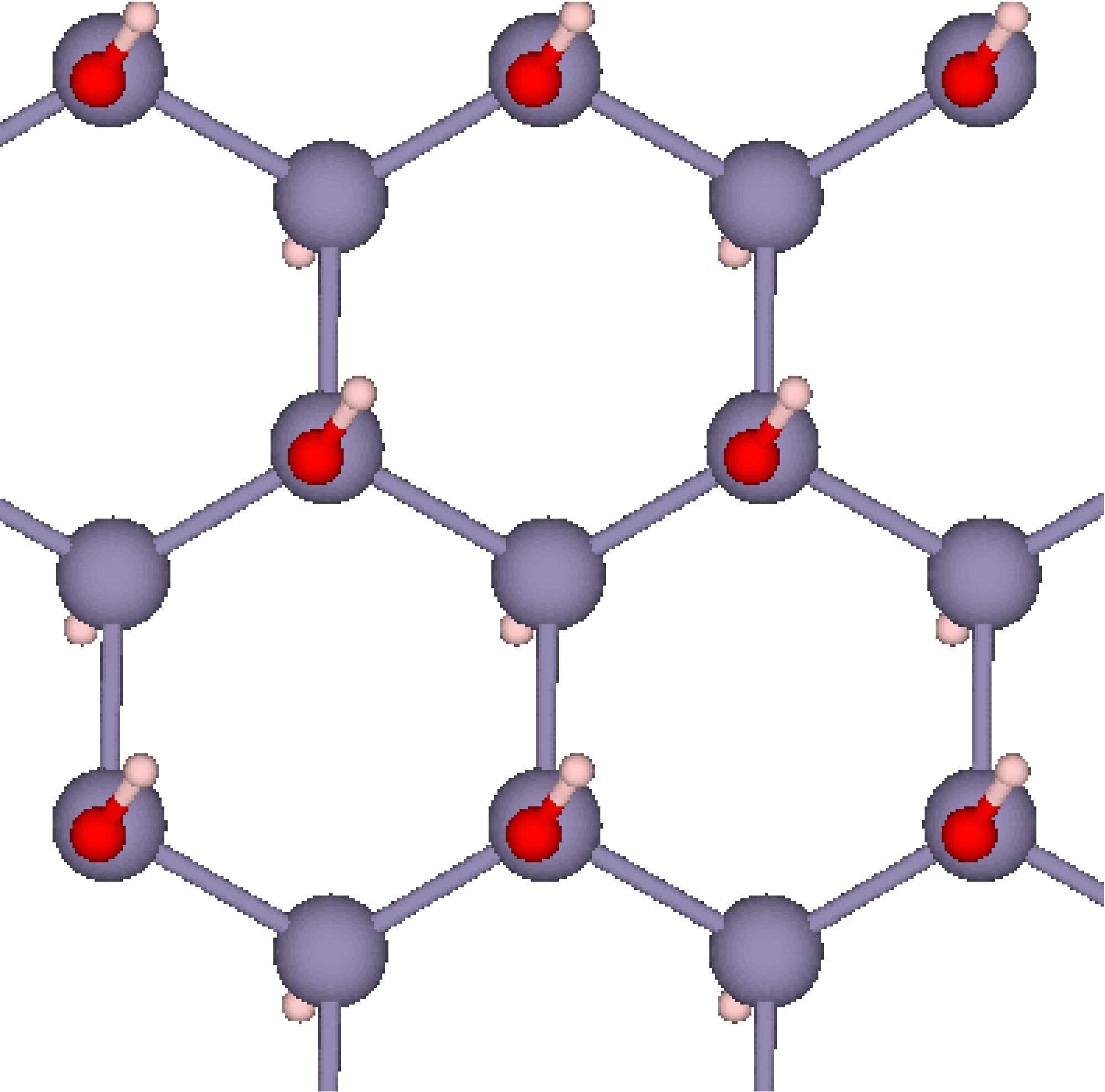}\hfill
  \includegraphics[width=4cm]{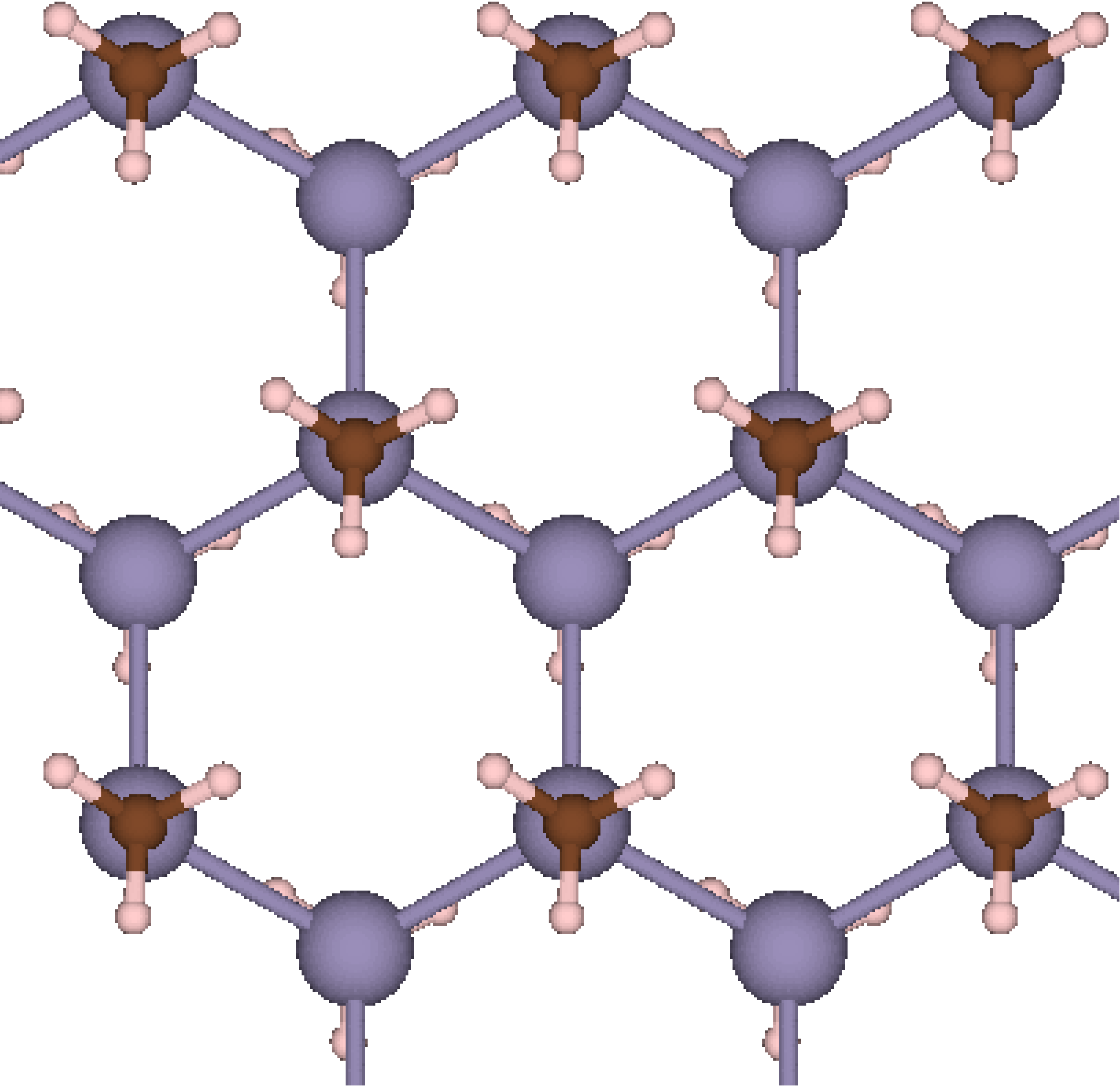}\\
  \includegraphics[width=4cm]{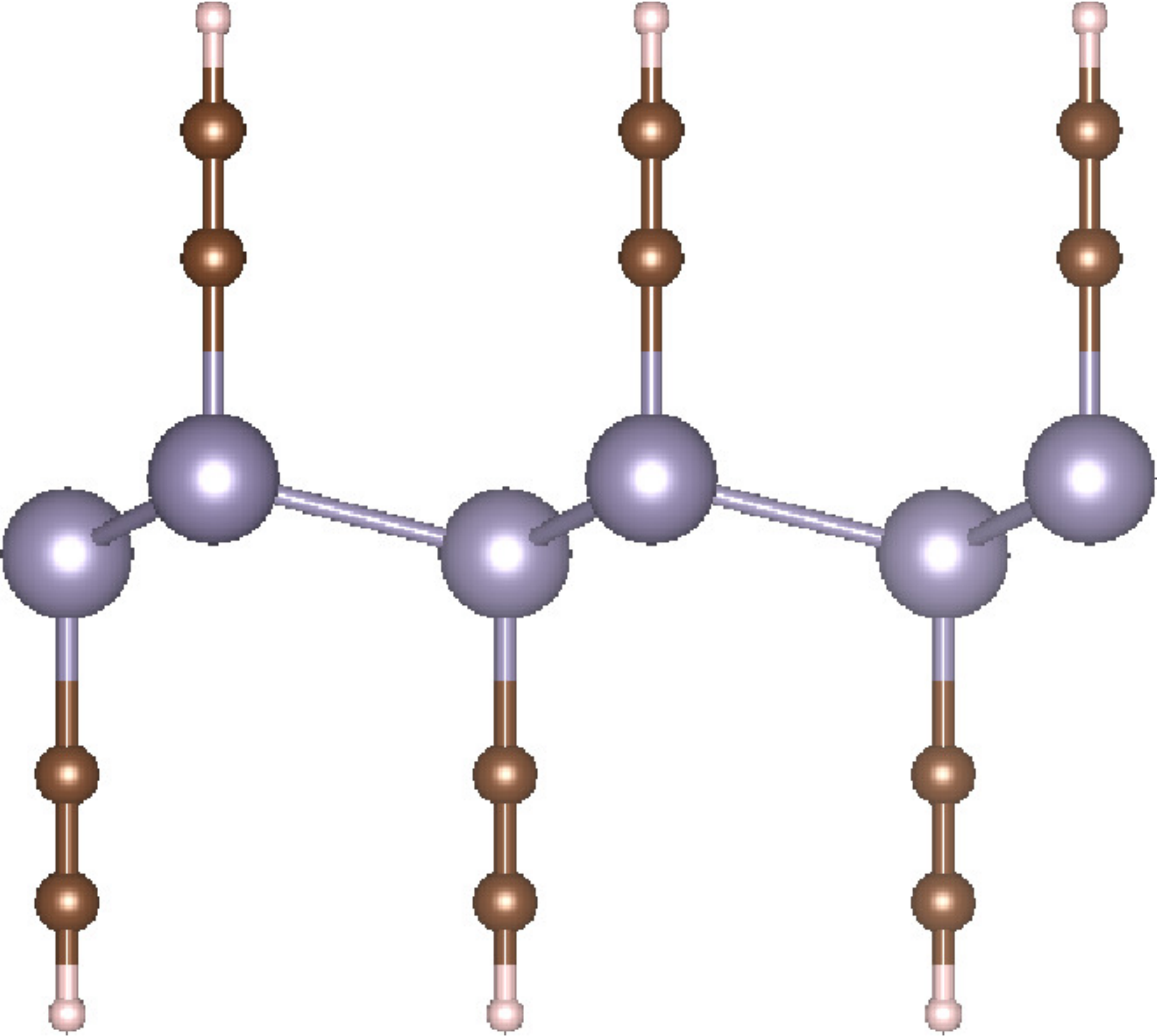}\hfill
  \includegraphics[width=4cm]{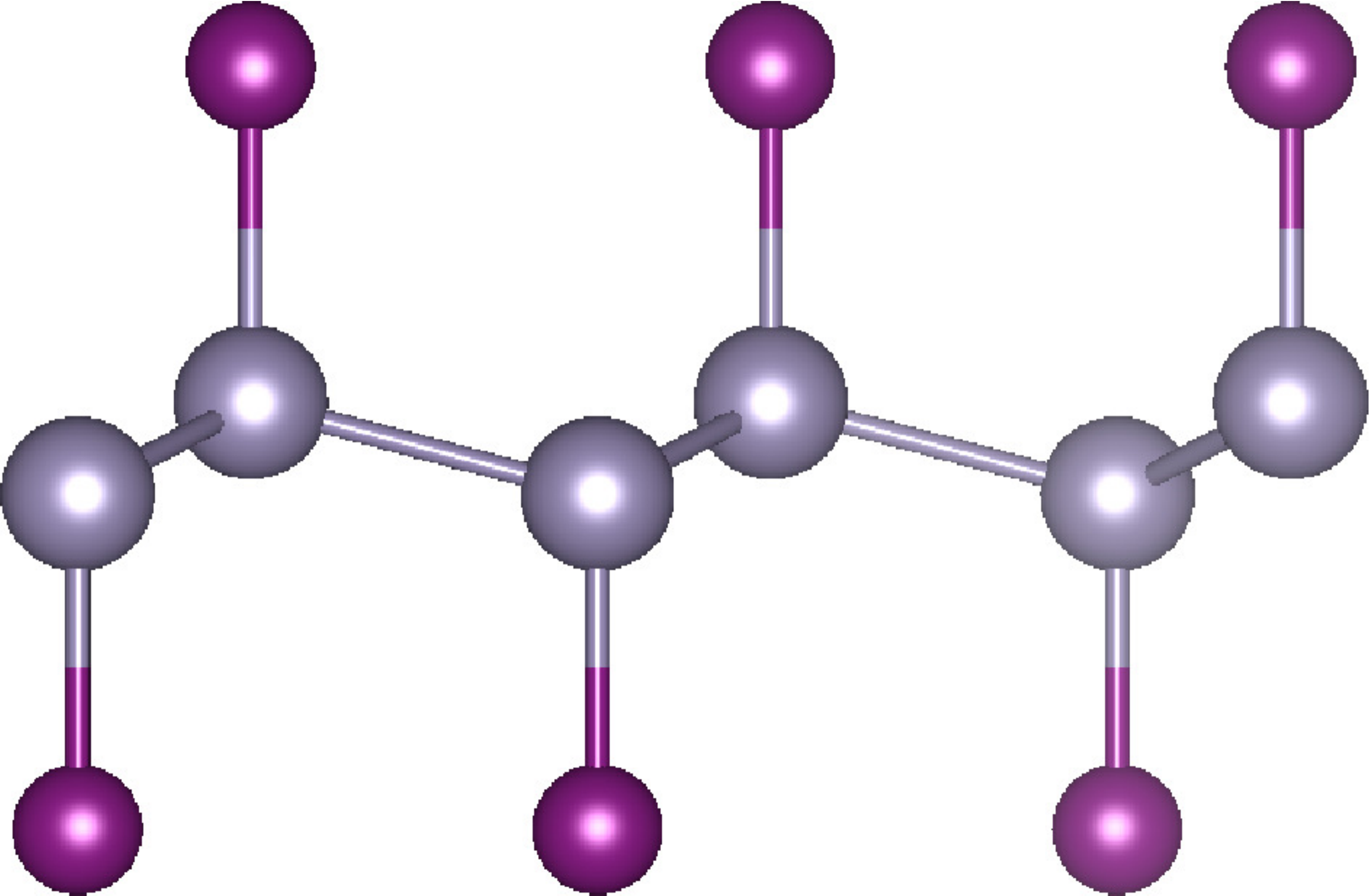}\hfill
  \includegraphics[width=4cm]{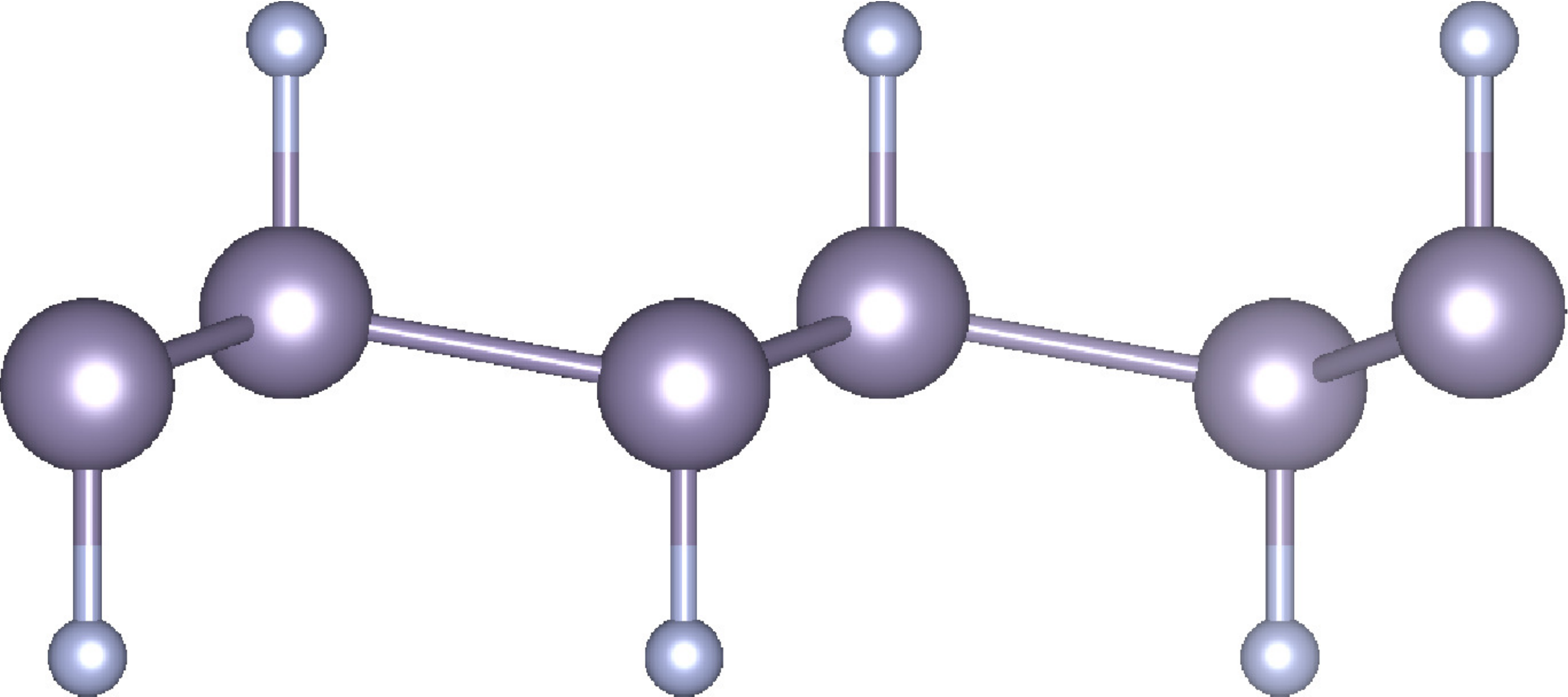}\\
  \includegraphics[width=4cm]{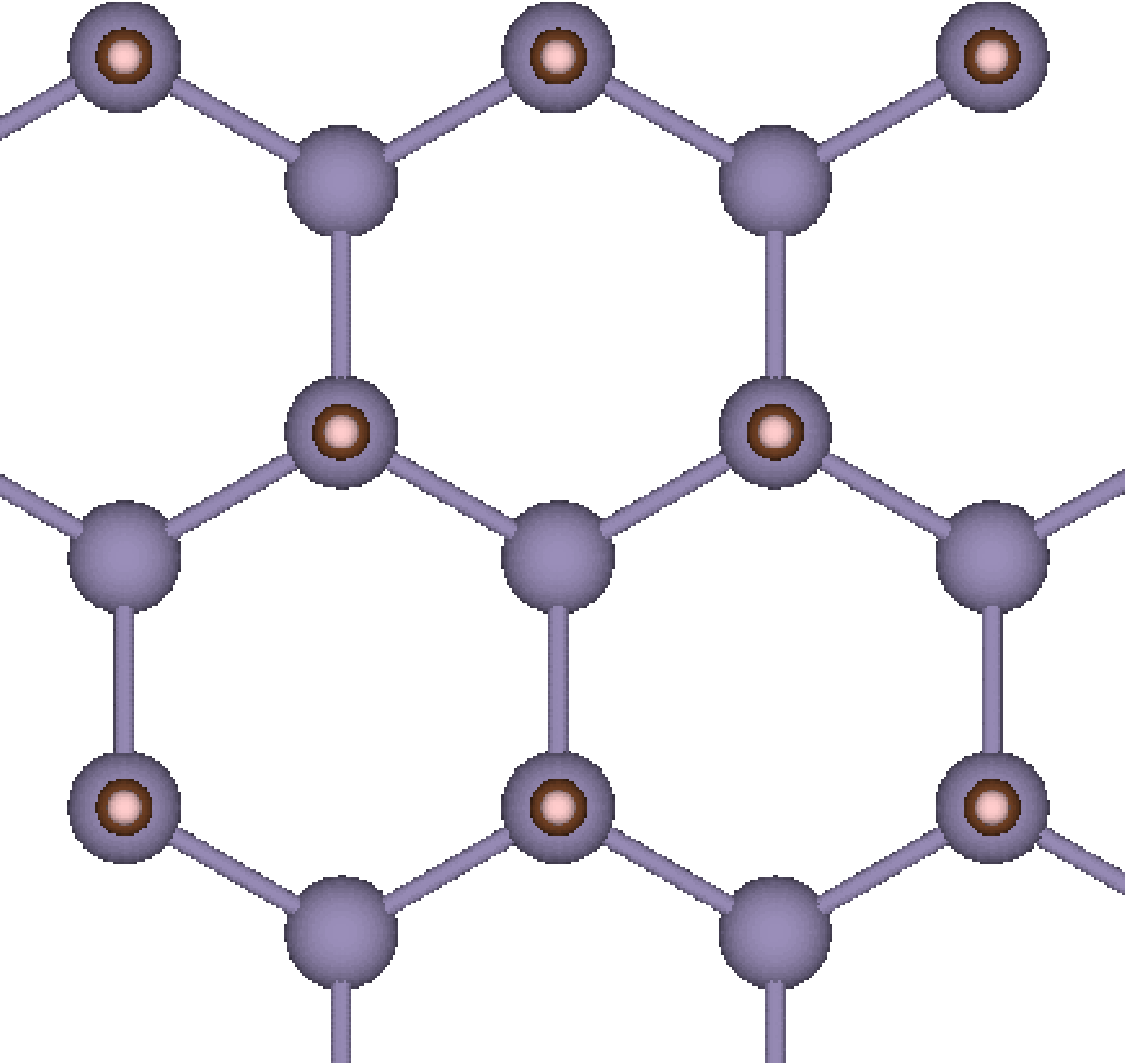}\hfill
  \includegraphics[width=4cm]{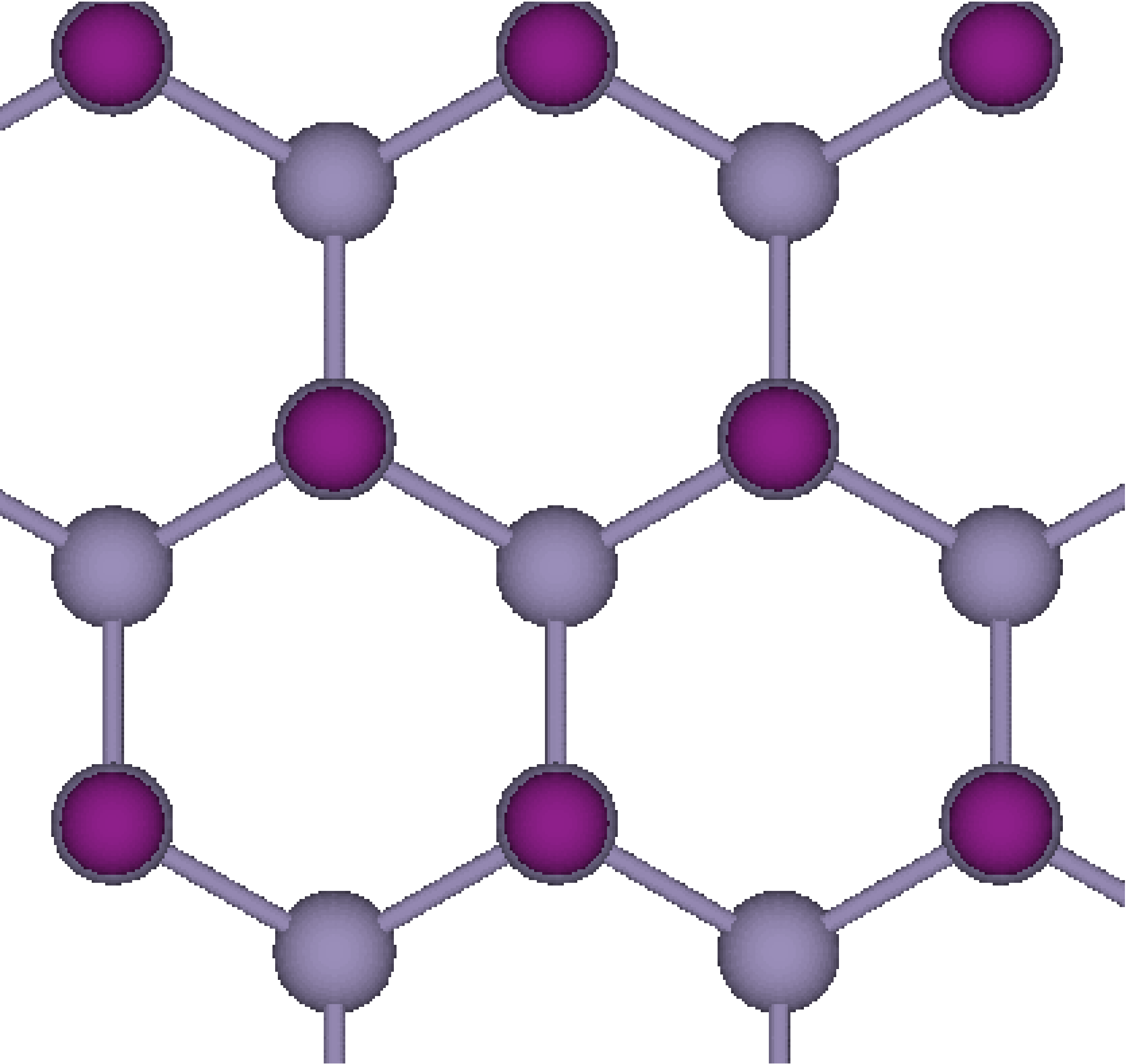}\hfill
  \includegraphics[width=4cm]{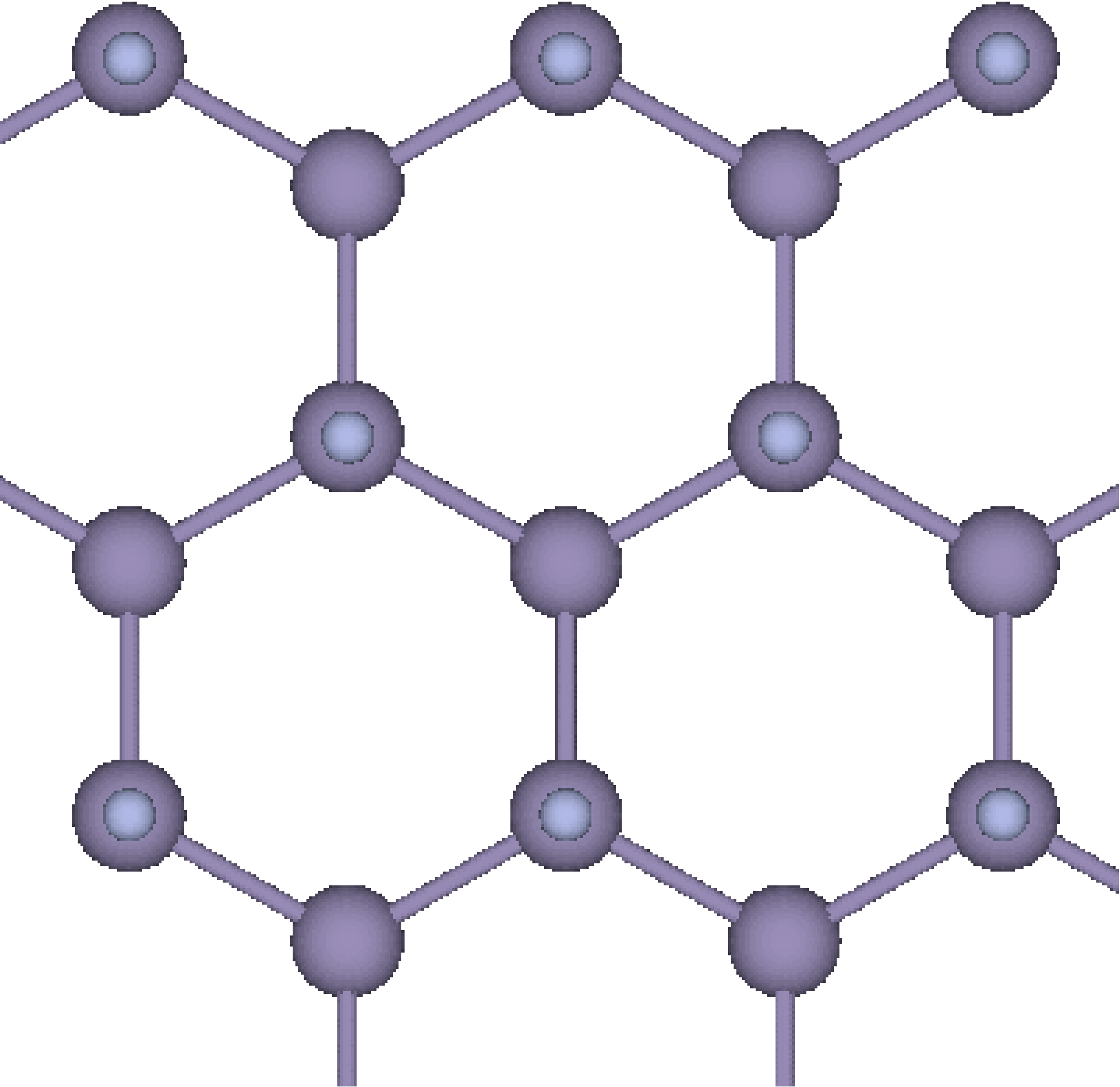}\\
    \caption{\label{fig:structures} Side view of relaxed
      tin layers adsorbed with organic ligands at one monolayer
      regime: a) -H, b) -OH, c) -CH$_3$, d) C$_2$H, e) -I and f)
      -F. Red, brown, blue, white and magenta are oxygen, carbon,
      hydrogen and germanium atoms, respectively.}
\end{figure}

The structural stability of the buckled and planar structure has been
investigated by varying the in-plane lattice parameter $a$ and fully
relaxing all the atoms in the unit cell. The relaxed energetically
most stable bare buckled structure is shown in
Fig.\,\ref{fig:structures}(a) (side view) and
Fig.\,\ref{fig:structures}(c) (top view). The optimized in-plane
lattice parameter for buckled (planar) stanene is $a$ = 4.67{\AA}
(4.85{\AA}). The buckled structure is 0.31\,eV energetically more
stable than the planar one. The Sn-Sn distance is 3.01\,{\AA}, as
shown in Table \ref{table:properties}. The calculated band structure of
buckled stanene is shown in Fig. \ref{fig:structuresbare} (b) and
density-of-states (DOS) os shown in Fig. \ref{fig:structuresbare} (d).
Stanene has a metallic character with small gap.

Calculations for the following ligands -H, -OH, -CH$_3$, -C$_2$H, -I
and -F have been performed. We have analized their relaxation, charge
density distribution, band structure and dielectric function. The
relaxed structures of funcionalized stanene are shown in
Fig.\,\ref{fig:structures}.

\begin{table*}[ht!]
\begin{tabular*}{1.0\textwidth}{@{\extracolsep{\fill}}lccccc}
\hline
ligand  & $a$(\AA)  & d$_{Sn-Sn}$ (\AA) & $d_{X}$(\AA) & $\Delta$(\AA) & E$_{gap}$(eV) \\
\hline
        &          &                  &              &               &       PBE         \\
\hline
-H        &  4.71   & 2.86             & 1.74         &  0.79         &    0.26        \\ 
-OH       &  4.89   & 2.88             & 2.02         &  0.66         &    0.22        \\
-CH$_3$   &  4.73   & 2.88             & 2.21         &  0.86         &    0.16        \\ 
-C$_2$H   &  4.84   & 2.92             & 2.12         &  0.67         &    0.22         \\ 
-I        &  4.90   & 2.83             & 2.73         &  0.86         &    0.34          \\ 
-F        &  4.99   & 2.93             & 1.96         &  0.53         &    0.28       \\
\hline
\end{tabular*}
\caption{Lattice parameter $a$, in-plane Sn-Sn distance $d_{Sn-Sn}$, Sn-X distance $d_{Sn-X}$, buckling $\Delta$, band gap $E_{gap}$ and formation energy $E_{f}$ of functionalized stanene.}
\label{table:properties}
\end{table*}

Upon adsorption of ligands, the initially buckled geometry of stanene
has a slight different buckling compared to the bare stanene, as shown
in Table\,\ref{table:properties}.Hybrid stanene structures have a
similar behavior to hybrid germanene where a small buckling is
reported\,\cite{arxiv2020,Nanomaterials2018,Zhao2016,PRB2014}.  This
is rather different from what has been found for similar organic
ligands on group-V nanostructures which become planar upon
adsorption\,\cite{JPCC2020,Kou2018a}. As a general feature, the
intralayer lattice parameter $a$ and consequently Sn-Sn distance are
sligthly dependent on the ligand type. On the other hand, the
ligand-stanene interaction should play an important role on the
stabilization of the hybrid structures.

In Fig. \ref{fig:structures} (a) is the Sn-H distance is 1.74 {\AA}
with Sn-Sn bond length $d$ of 2.86({\AA}). The Sn-OH structure is
shown in Fig. \ref{fig:structures} (b) and has $a$ = 2.02\,{\AA} and
Sn-Sn distance $d$ = 2.88 {\AA}. The ${\rm Sn-CH_3}$ structure shown
in Fig. \ref{fig:structures} (c) has $a$ = 2.21\,{\AA} and Sn-Sn
distance $d$ of 2.88\,\AA. The in-plane lattice constant is slightly
lower for ${\rm Sn-C_2H}$ shown in Fig.\,\ref{fig:structures} (d)
(2.12 {\AA}).  In Fig.\,\ref{fig:structures} (e) the functionalization
with I is seen. It has a Sn-I distance of 2.73\,\AA and Sn-Sn bond
lengths of 2.83\,\AA.  Finally for Sn-F one has Sn-F equal to 1.96\,AA
and Sn-Sn of 2.93 \AA. As the ligands we consider here are small,
ligand-Sn interaction should play a major role in stabilizing hybrid
Sn-ligand structures.

\section{Charge density difference}

\begin{figure*}[htpb!]
  \centering
  \includegraphics[width=4cm]{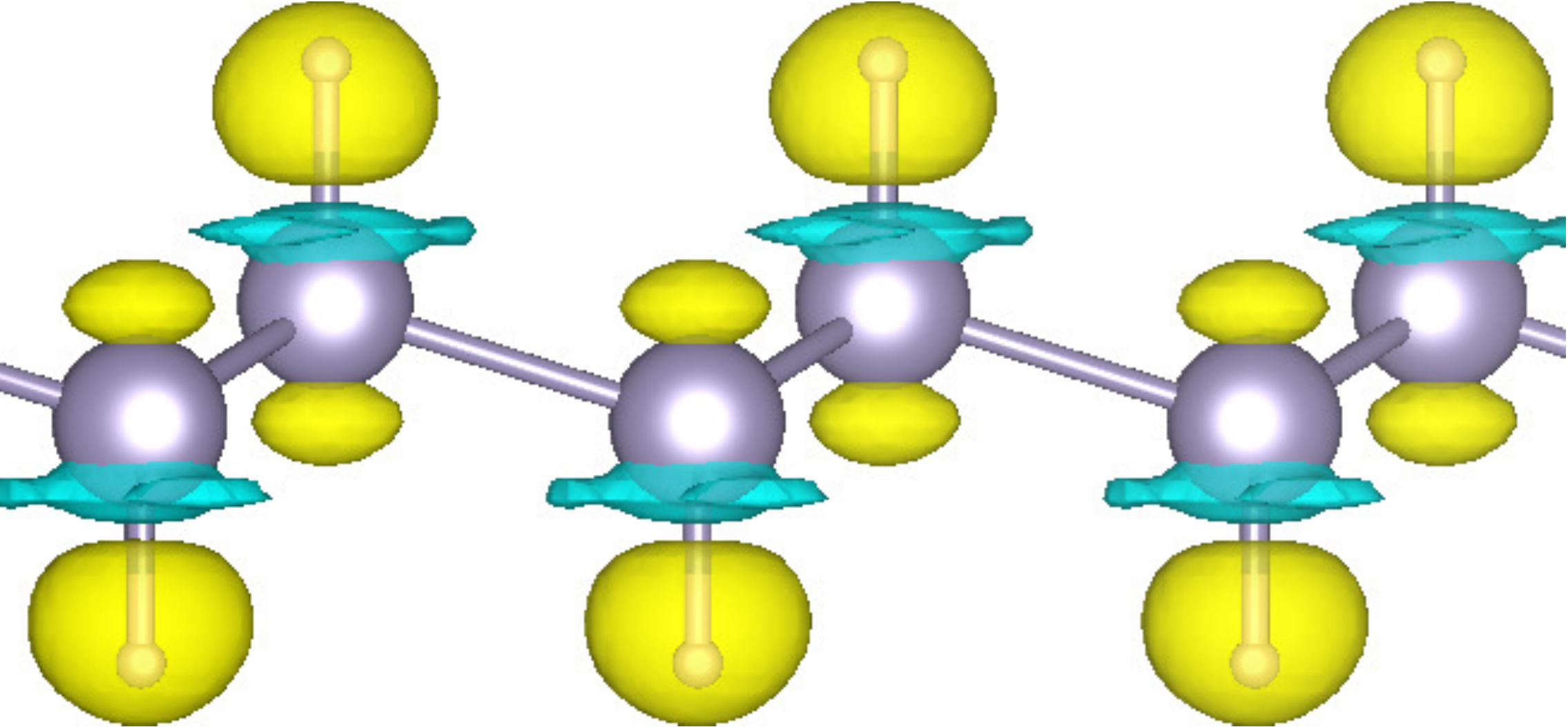}\hfill
  \includegraphics[width=6cm]{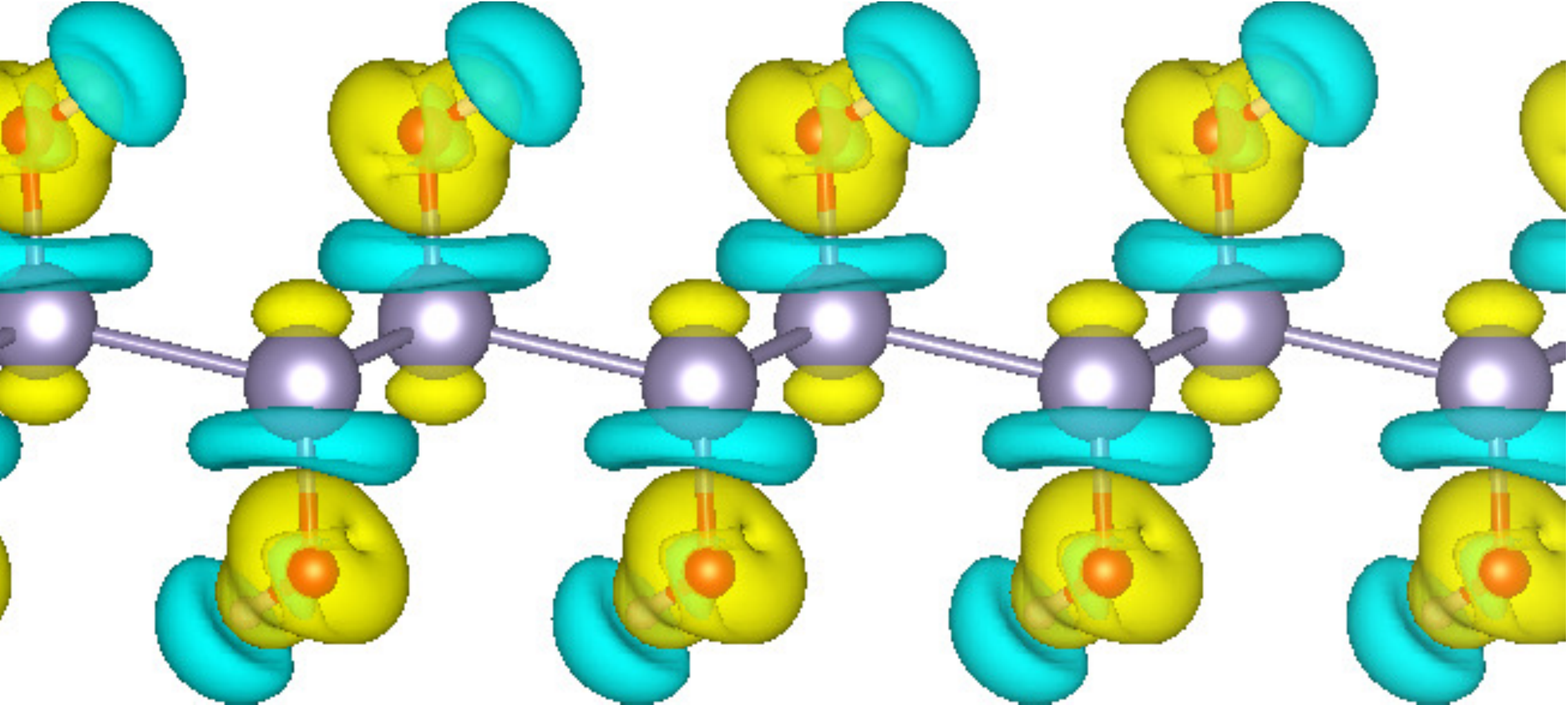}\hfill
  \includegraphics[width=5cm]{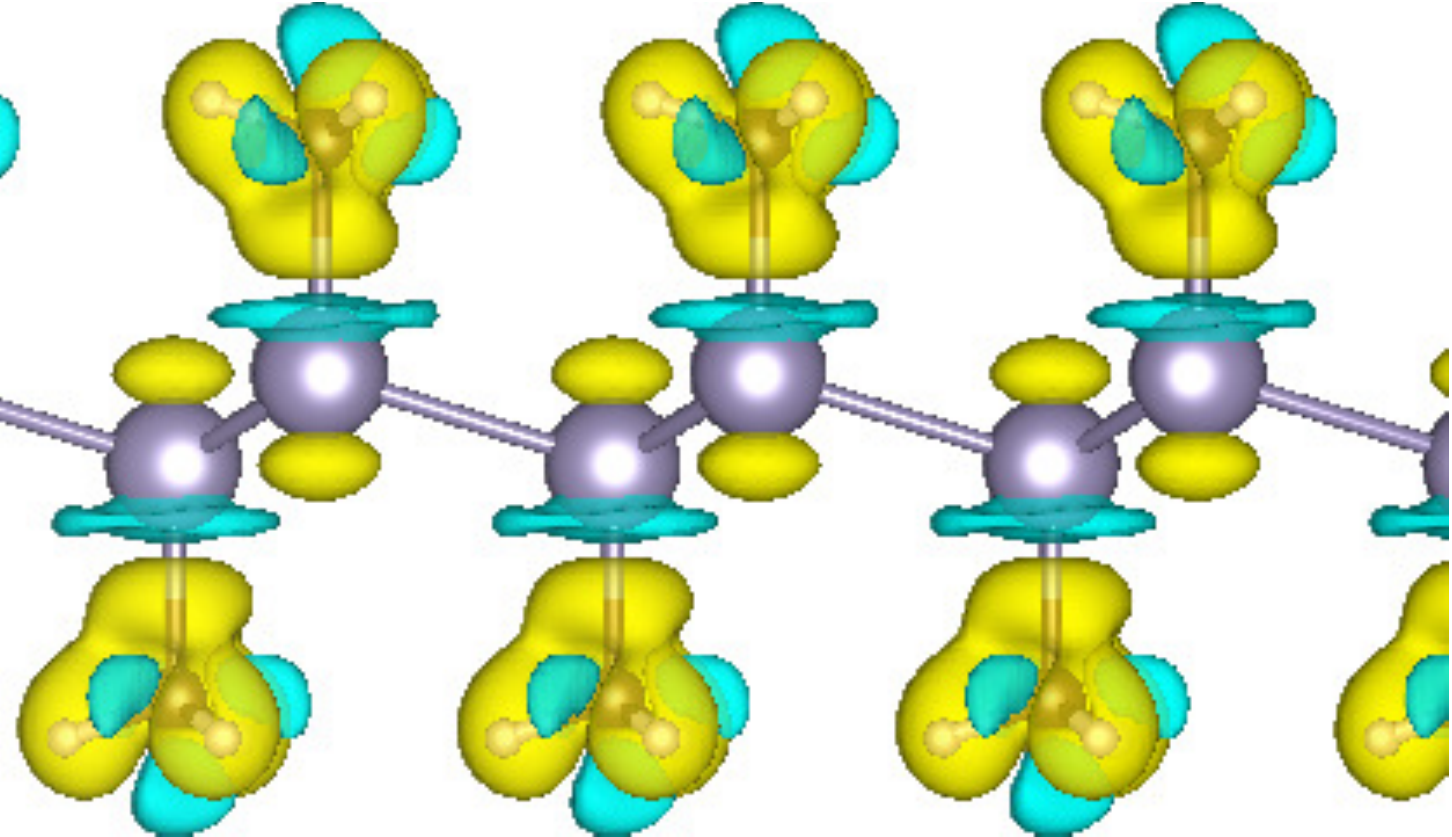}\\
  \includegraphics[width=5cm]{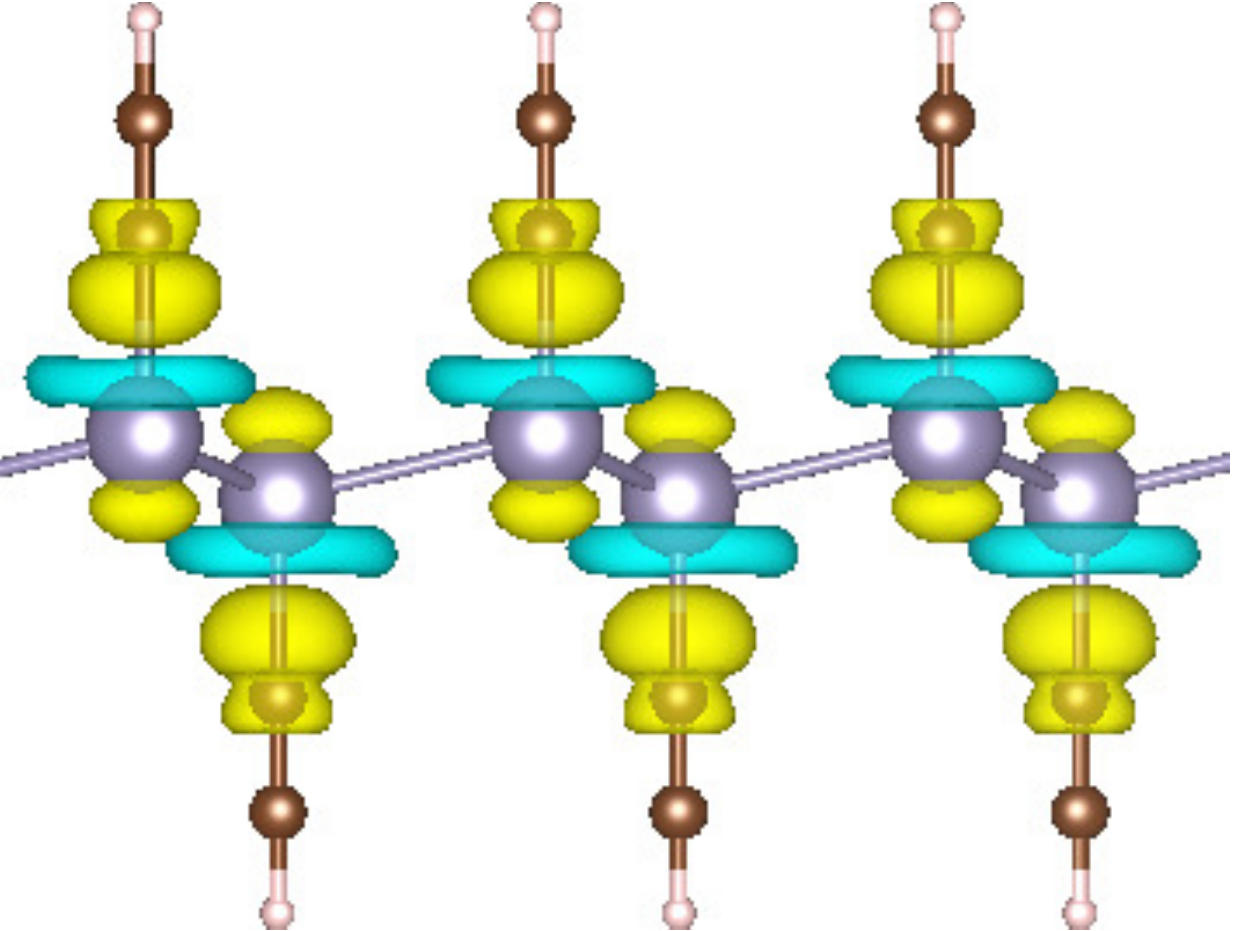}\hfill
  \includegraphics[width=6cm]{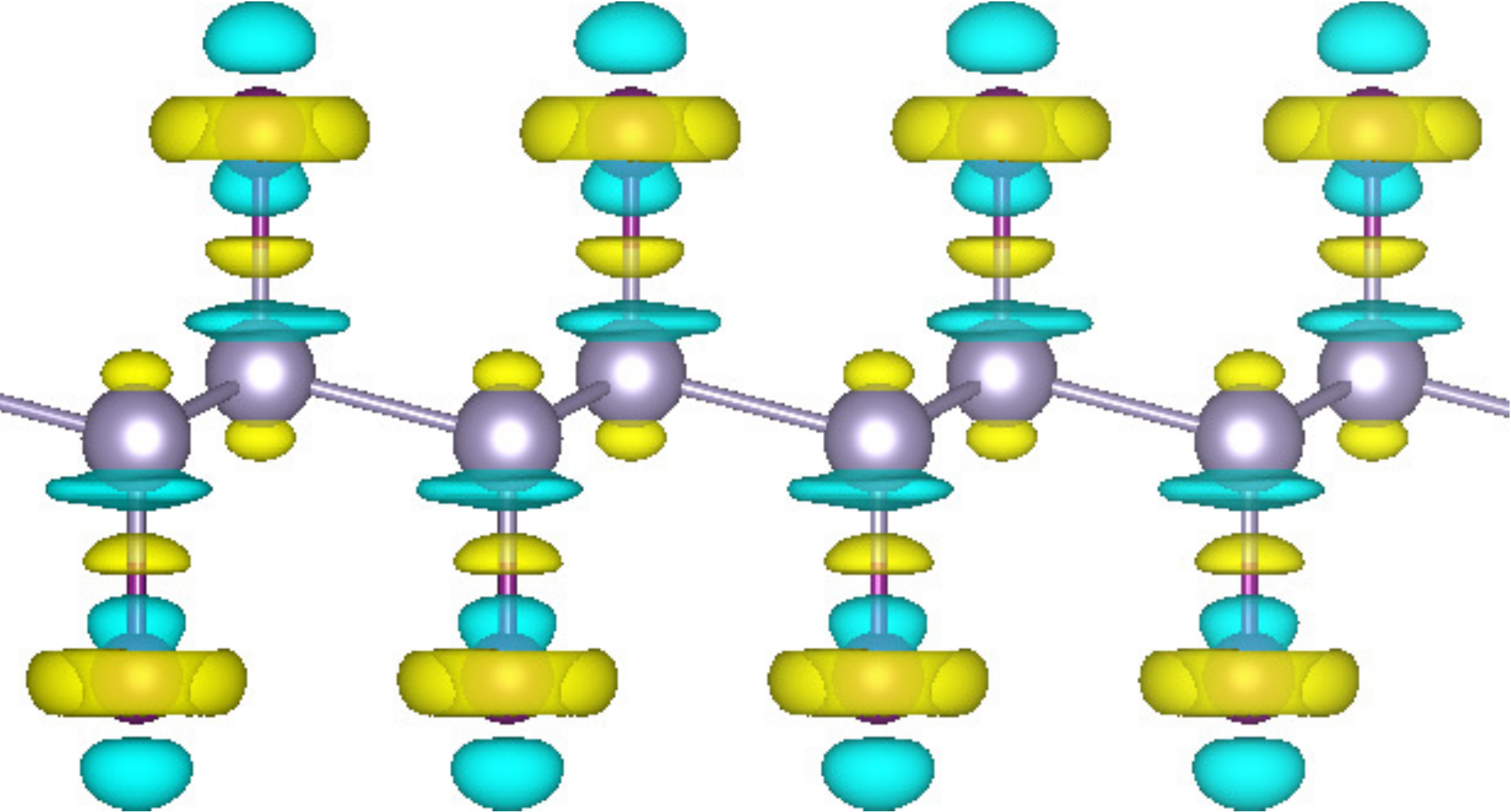}\hfill
  \includegraphics[width=5cm]{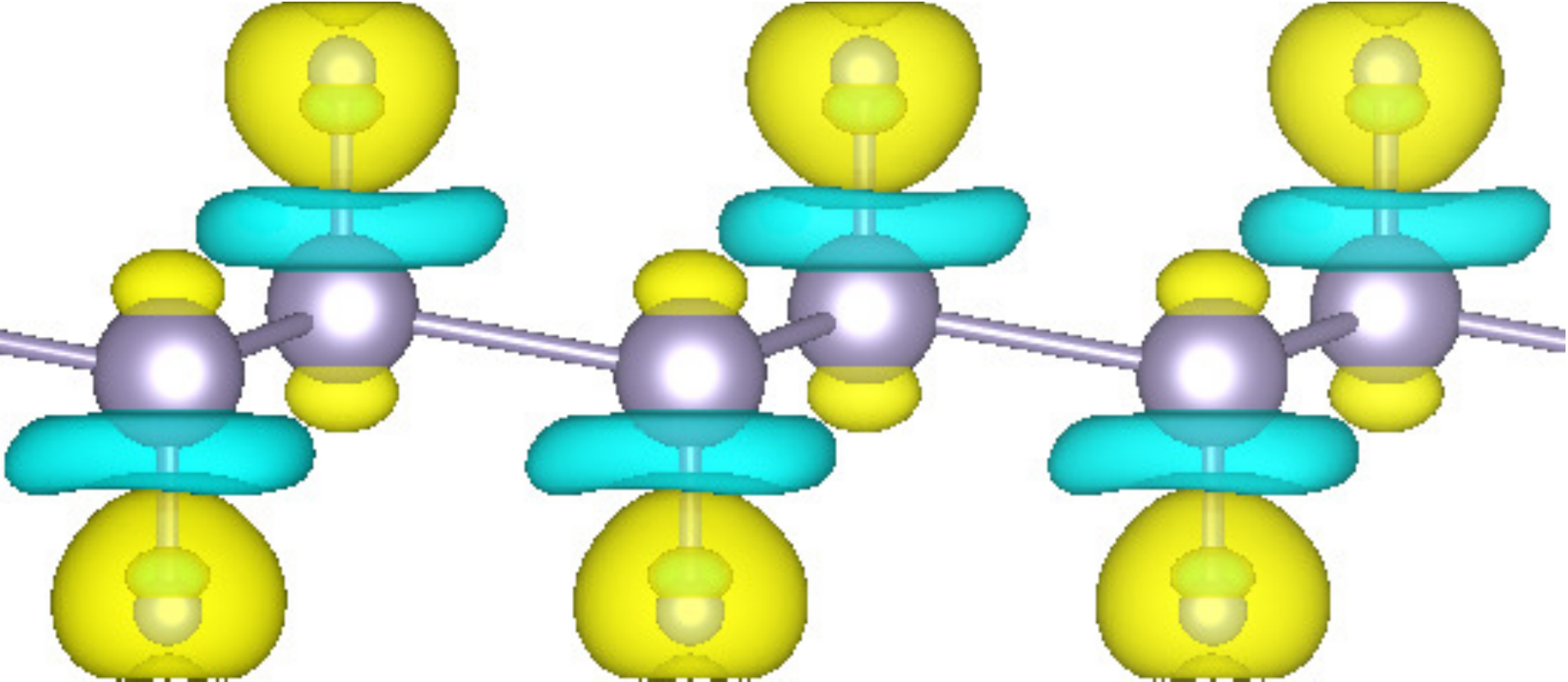}
\caption{\label{fig:chargediff} Charge density difference of modified
  stanene structures with a) -H, b) -OH, c) -CH$_3$, d)
  C$_2$H, e) -I and f) -F. Blue indicates a loss of electrons
  whereas yellow indicates accumulation of electrons. Isosurface
  levels are set to 0.002 $e$/\AA$^3$.}
\end{figure*}

The nature of the chemical bonds in the hybrid systems was
investigated by calculating the charge density difference between the
electronic densities of the adsorbed stanene and their constituent
systems. The electronic density difference shown in
Fig.\,\ref{fig:chargediff} is given by ${\rm \Delta\rho = \rho^{Sn-X}
  -\rho^{Sn} - \rho^{-X}}$, where ${\rm \rho^{Sn-X}}$ is the
electronic density of the hybrid Sn-X layers, $\rho^{Sn}$ and
${\rho^{-X}}$ are the charge densities of stanene and the
ligand, respectively, calculated at fixed atomic positions after
structure optimization of the hybrid layers. As a general feature,
charge accumulation on the ligand (yellow) and charge withdrawal
(blue) close to the tin atom is seen.

Sn-H hybrid layers shown in Fig.\,\ref{fig:chargediff}(a) have a
larger electron density at hydrogen atoms and a withdraw of electrons
at tin sites. Sn-OH also shows accumulation of charge at the ligand,
mainly on oxygen and carbon atoms, while charge is withdrawn at tin
sites as shown in Fig.\,\ref{fig:chargediff}(b). The complex Sn-CH$_3$
shows that electronic charge is accumulated the ligand upon
adsorption on stanene. Furthremore, between C and Sn bonds there
is an excess of electrons as it can be seen in
Fig.\,\ref{fig:chargediff}(c) . The Sn-${\rm C_2H}$ hybrid also shows
accumulation of charge at the ligand, mainly on the oxygen and carbon
atoms, while charge is depleted at the tin site, as shown in
Fig.\,\ref{fig:chargediff}(d). The complex Sn-I also shows
accumulation of electrons on the ligand oxygen and carbon atoms, while
regions close to the tin are depleted of electrons. At the tin site
and also between C and Sn bonds there is an excess of electrons as
seen in Fig.\,\ref{fig:chargediff}(e). The complex Sn-F also shows
accumulation of electrons on the ligand oxygen and carbon atoms, while
regions close to the tin are depleted of electrons. At the tin site
and also between C and Sn bonds there is an excess of electrons as
seen in Fig.\,\ref{fig:chargediff}(f).

One can therefore conclude that due to the charge
accumulation/withdraw in the modified tin layers, the reactivity of
the whole system changes. This feature could be quite useful for
further functionalization of organic or biomolecules. Taking into the
account the electronegativity trend in the periodic table, we have the
following order for electronegativity: F $>$ I and C $\approx$ I, as
calculated in Ref.\,\cite{cartagena98}.  These calculations allowed
determine the correlation between the electronegativity between Sn and
CH$_3$ and H groups.  According to these calculations, it is expected
that the electron affinity follows: Sn-F $>$ Sn-H $>$ Sn-CH$_3$. If we
consider our results for the adsorbed stanene, our results corroborate
with these trends.

\subsection{Electronic properties}

\begin{figure}[ht!]
\begin{center}
\begin{tabular}{cc}
\includegraphics[width = 7.5cm, scale=1, clip = true, keepaspectratio]{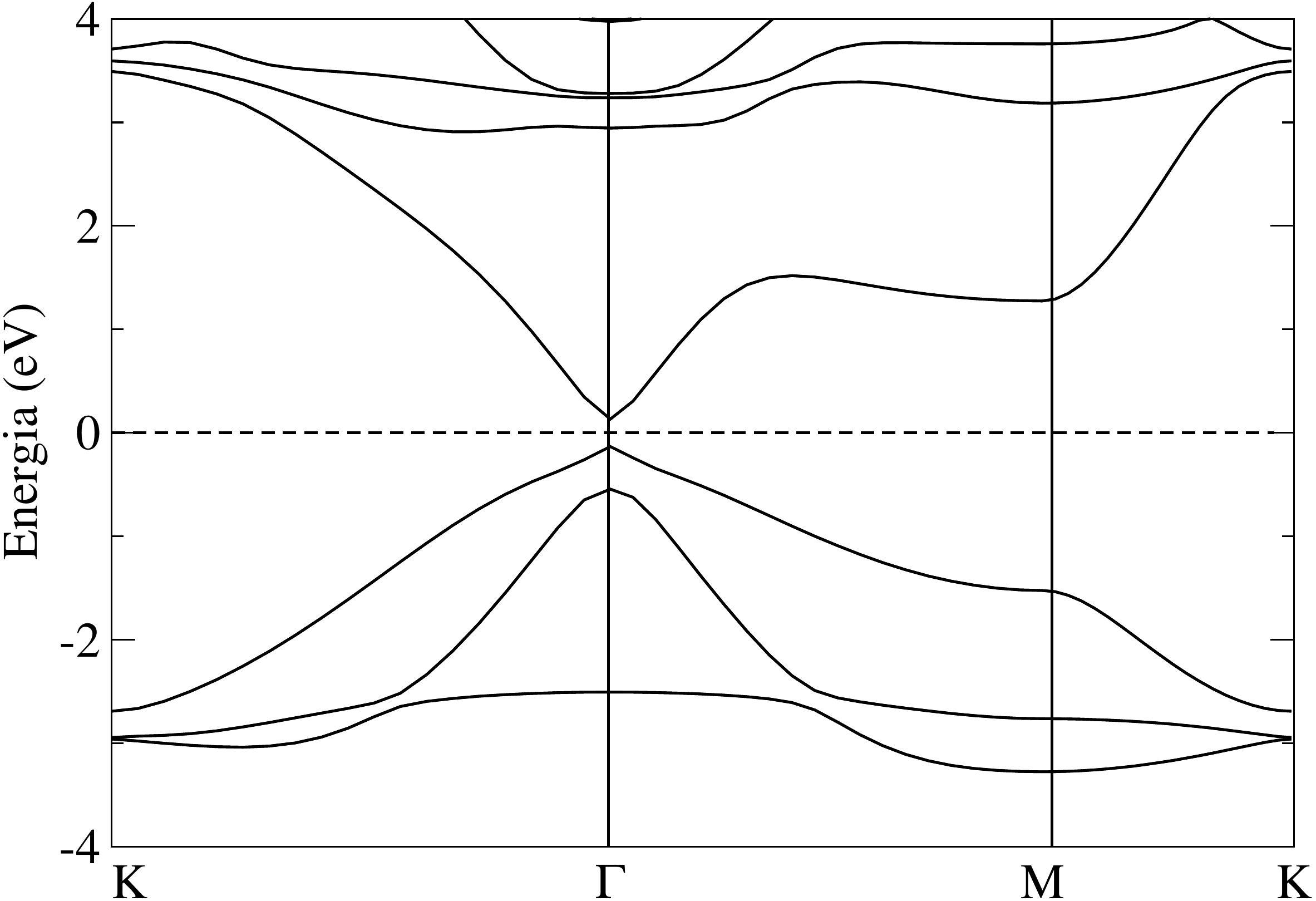}&
\includegraphics[width = 7.5cm, scale=1, clip = true, keepaspectratio]{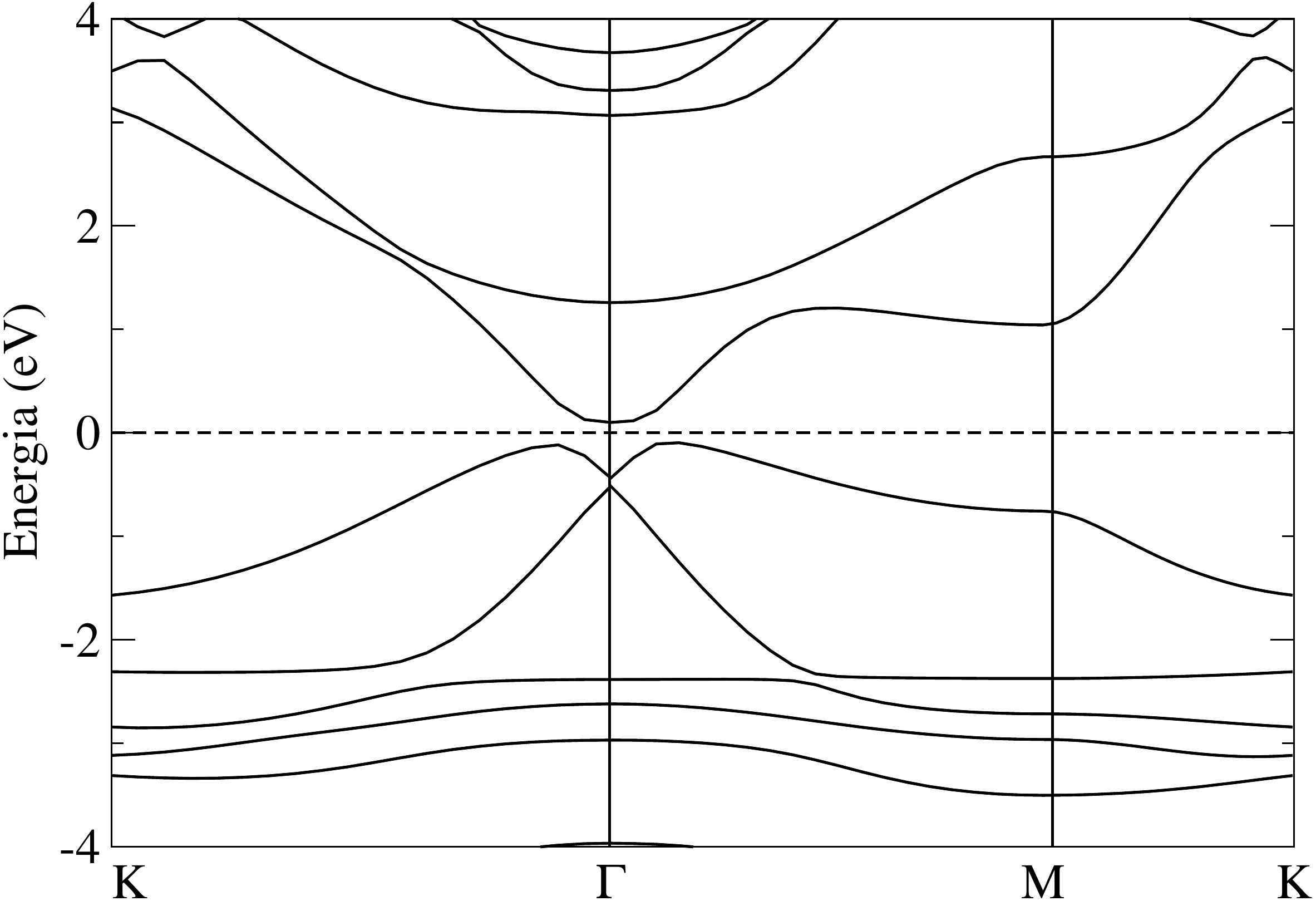}\\
\includegraphics[width = 7.5cm, scale=1, clip = true, keepaspectratio]{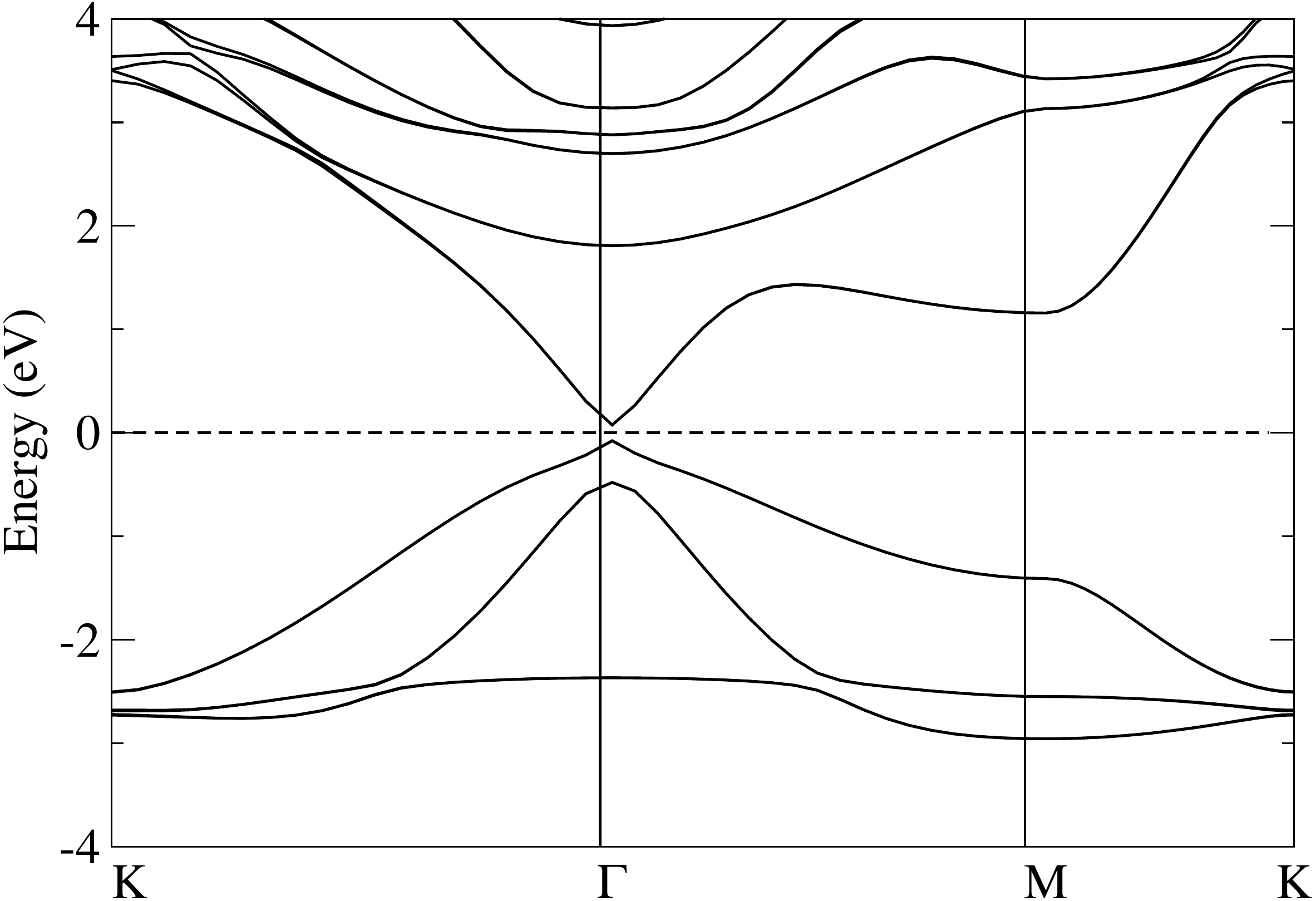}&
\includegraphics[width = 7.5cm, scale=1, clip = true, keepaspectratio]{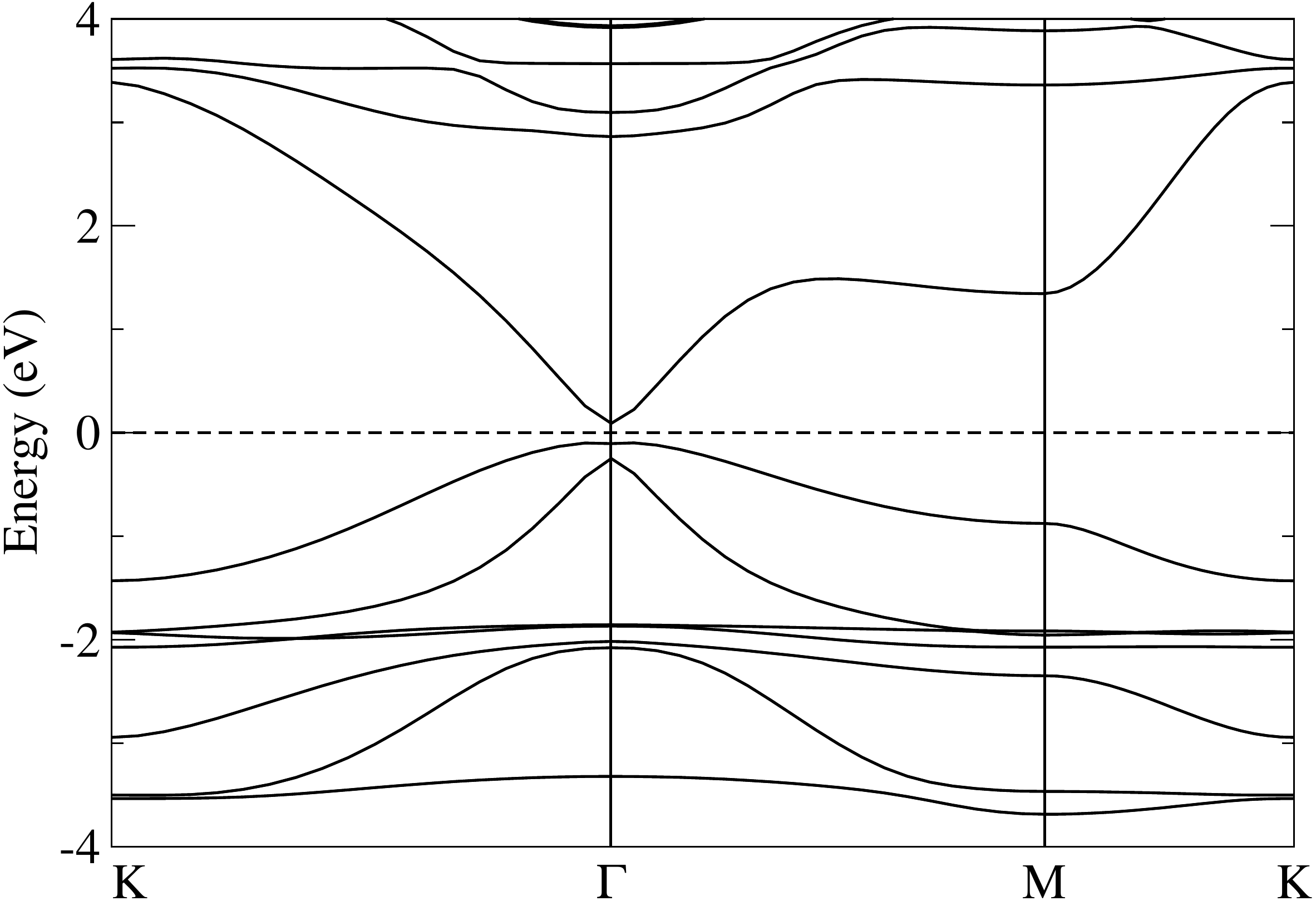}\\
\includegraphics[width = 7.5cm, scale=1, clip = true, keepaspectratio]{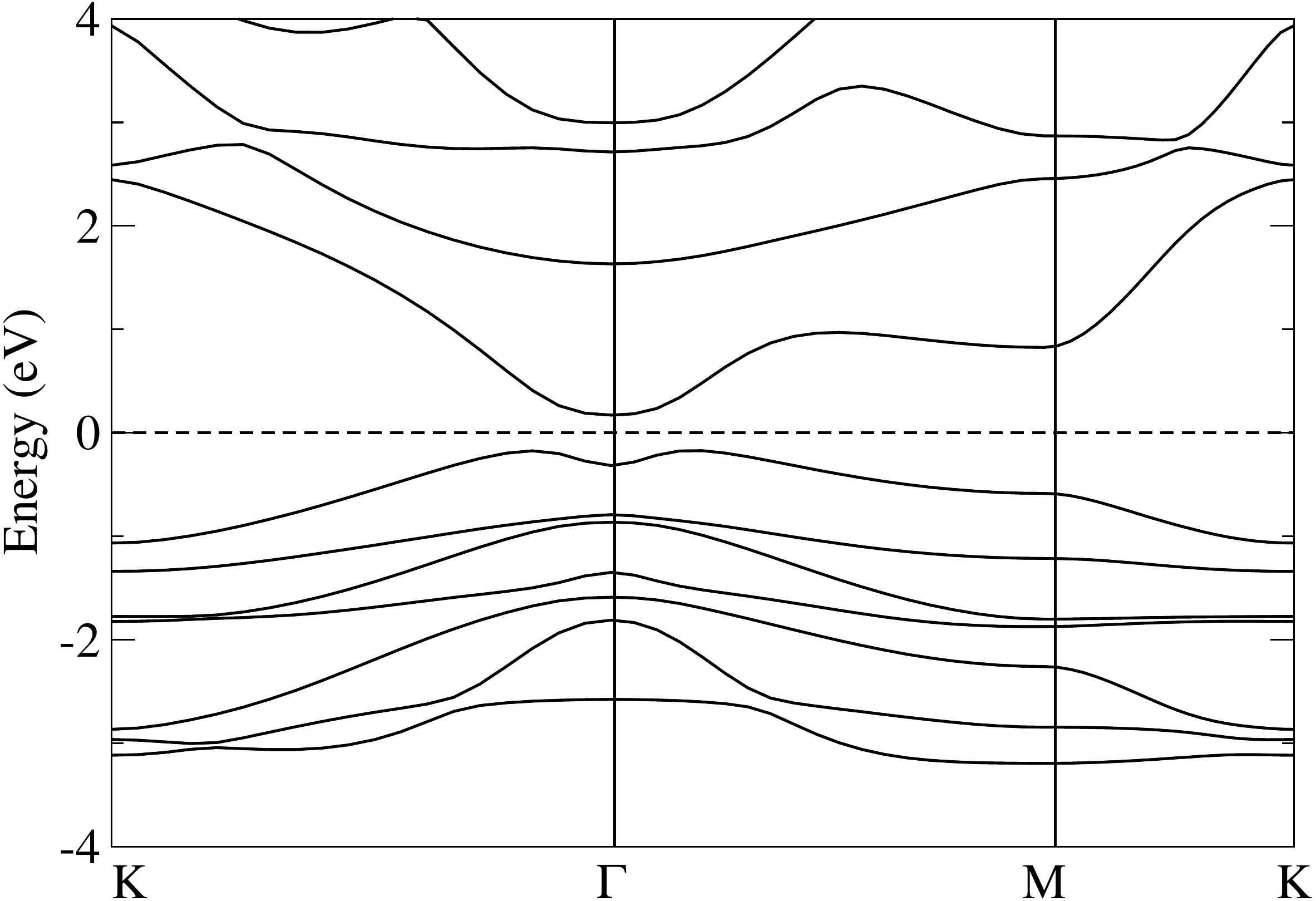}&
\includegraphics[width = 7.5cm, scale=1, clip = true, keepaspectratio]{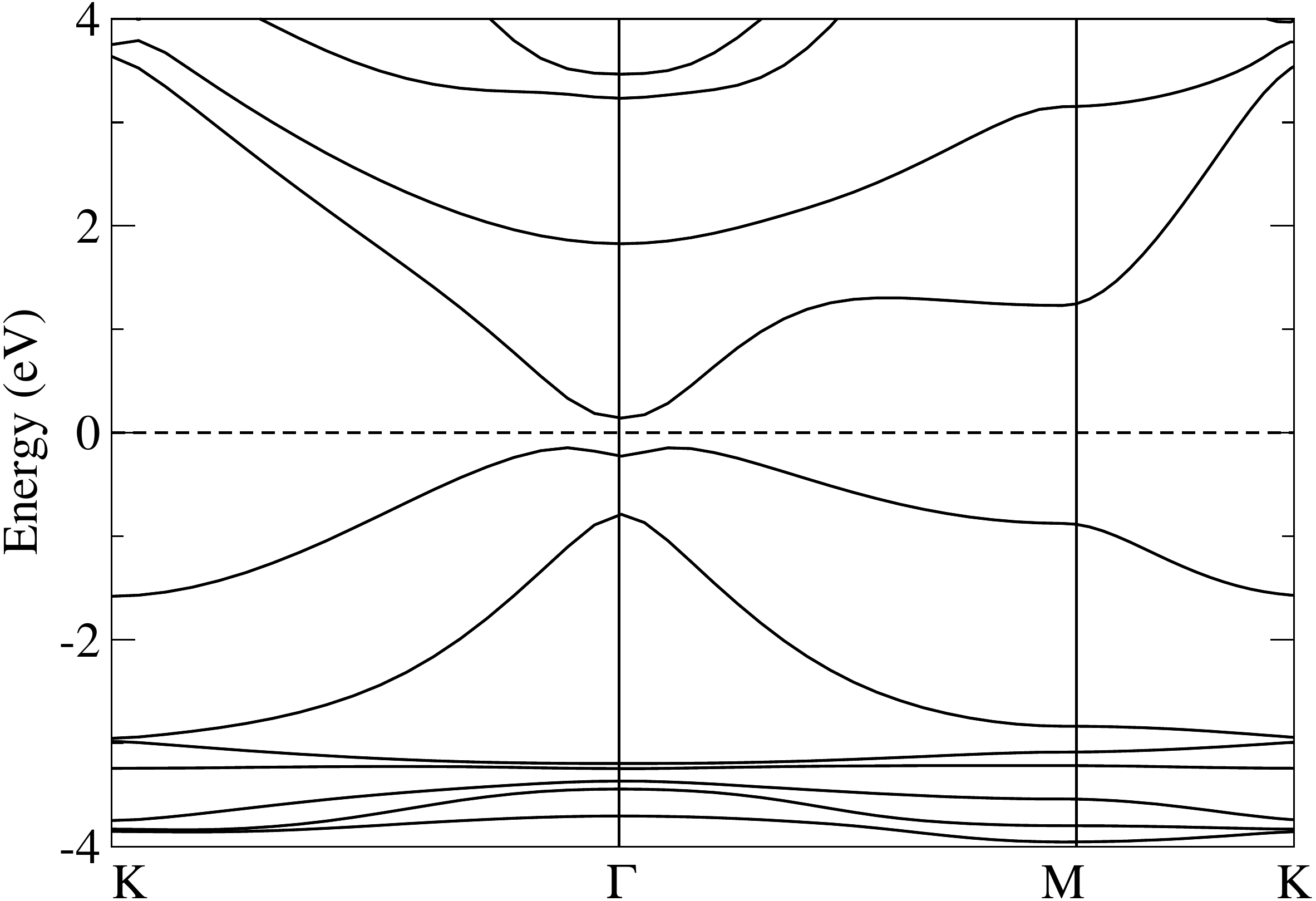}
\end{tabular}
\end{center}
\caption{Band structure of modified stanene structures. a) -H, b) -OH, c) -CH$_3$, d) -C$_2$H, e) -I and f) F. The zero of energy is set at zero.}
\label{fig:projected_bands}
\end{figure}

\begin{figure}[ht!]
\begin{center}
\begin{tabular}{cc}
\includegraphics[width = 7.5cm, scale=1, clip = true, keepaspectratio]{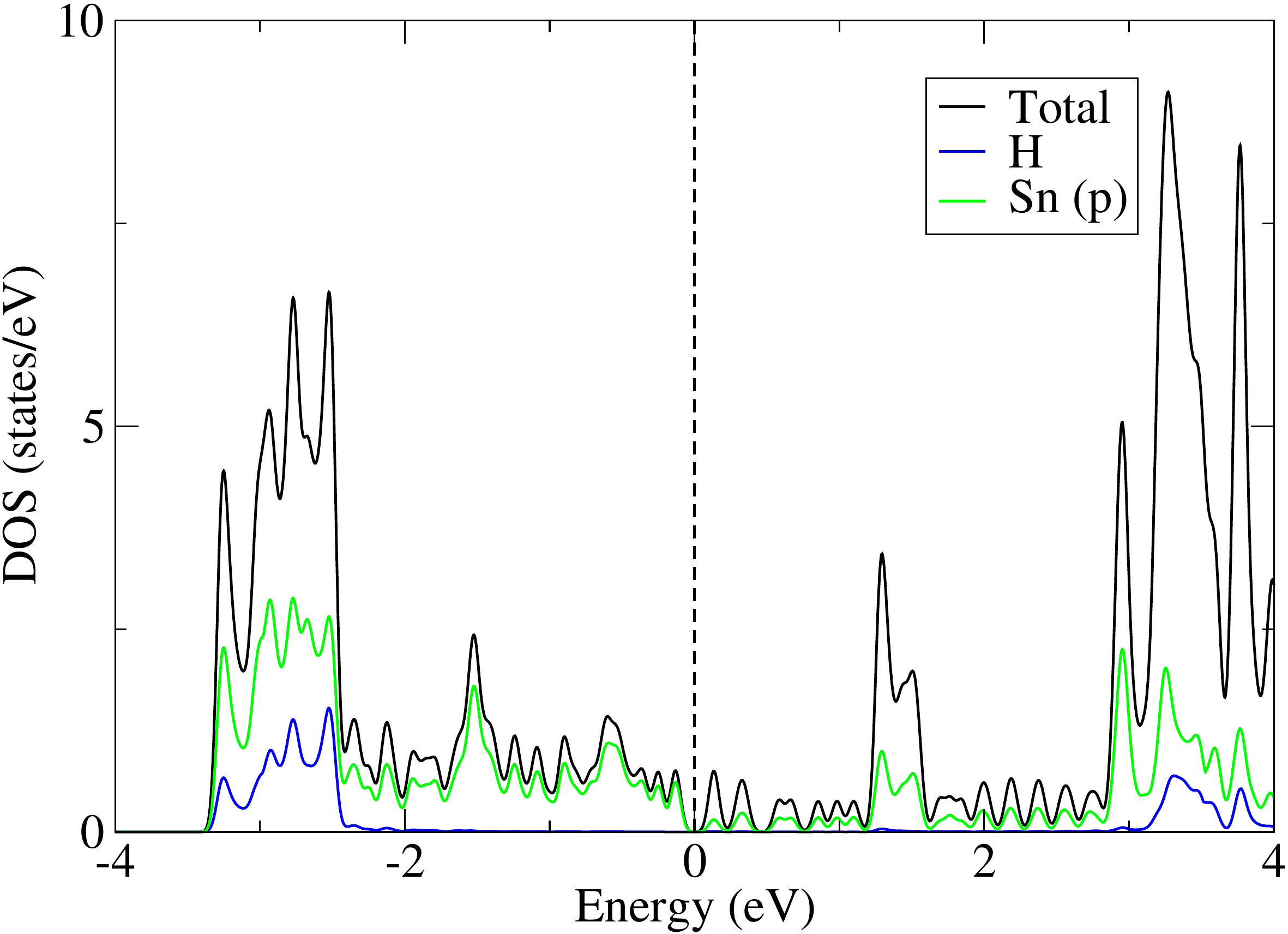}&
\includegraphics[width = 7.5cm, scale=1, clip = true, keepaspectratio]{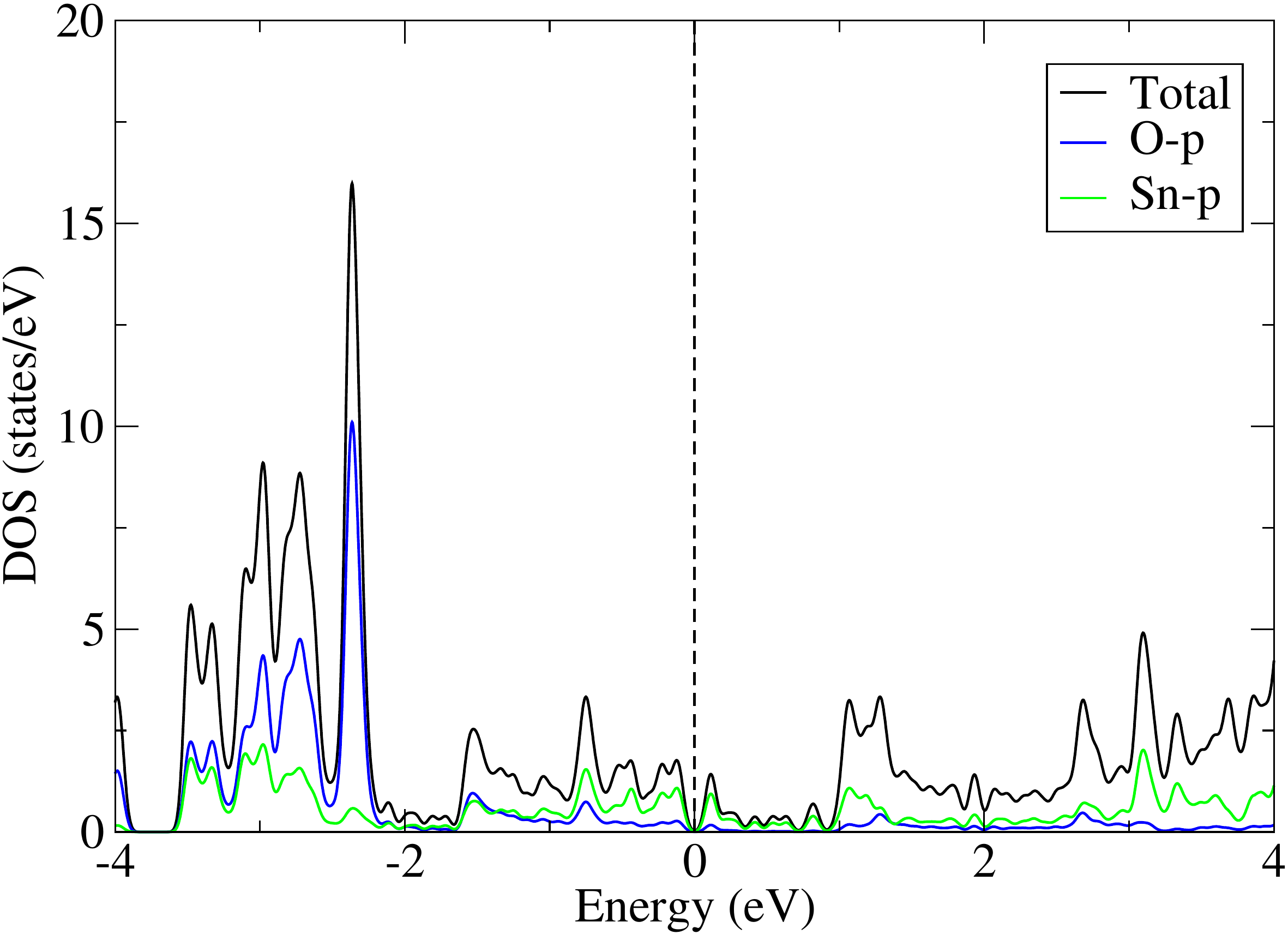}\\
\includegraphics[width = 7.5cm, scale=1, clip = true, keepaspectratio]{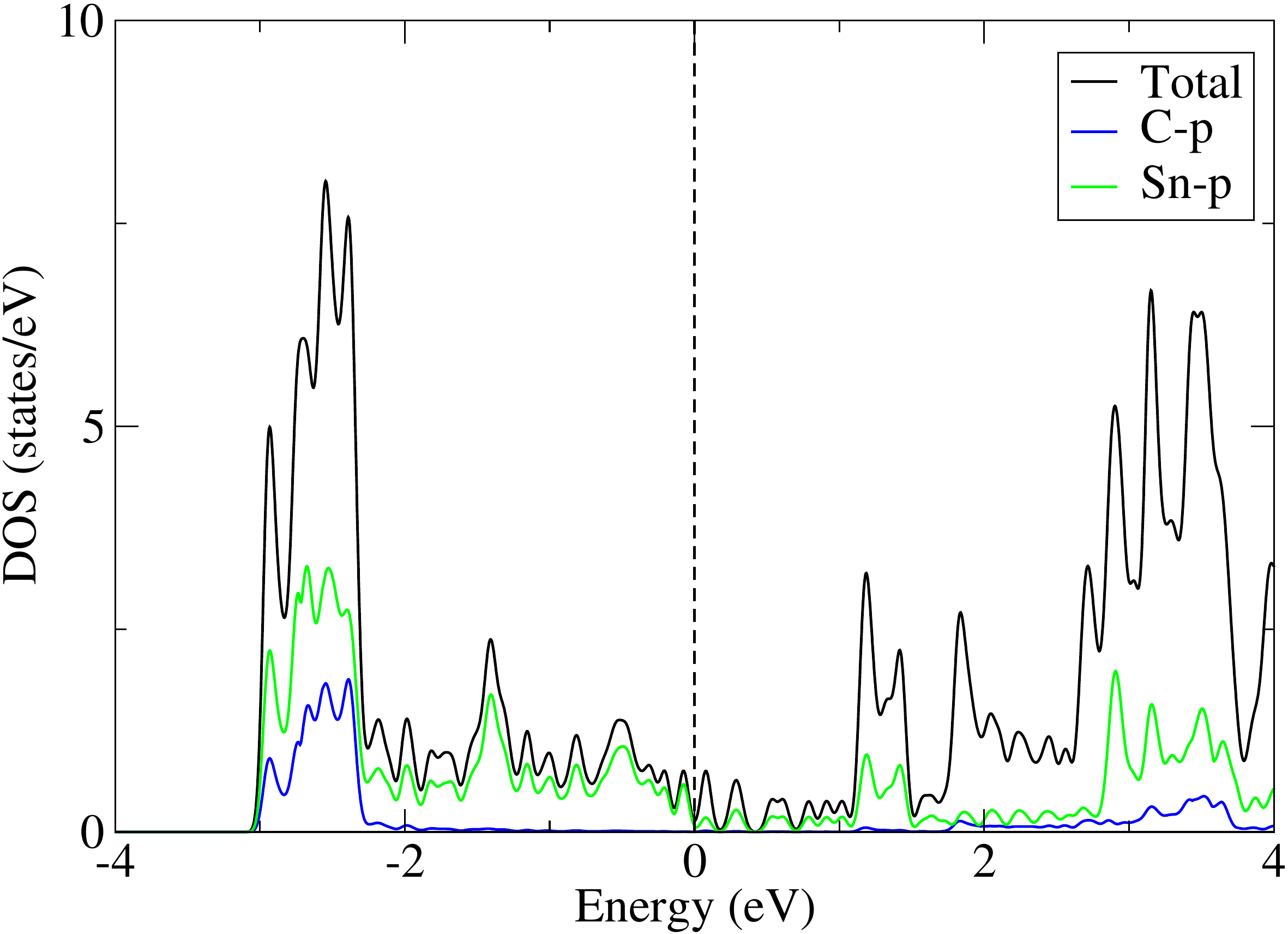}&
\includegraphics[width = 7.5cm, scale=1, clip = true, keepaspectratio]{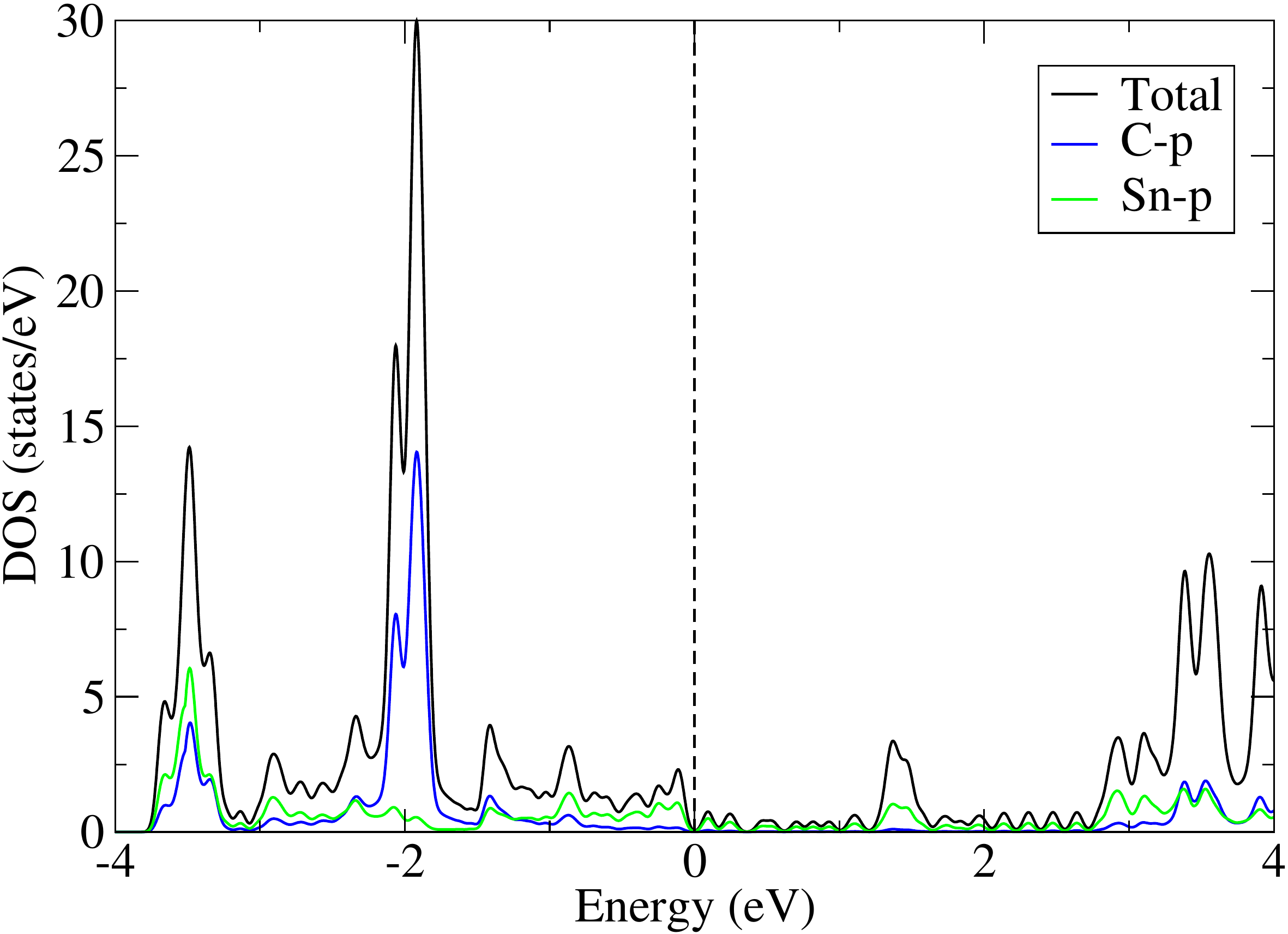}\\
\includegraphics[width = 7.5cm, scale=1, clip = true, keepaspectratio]{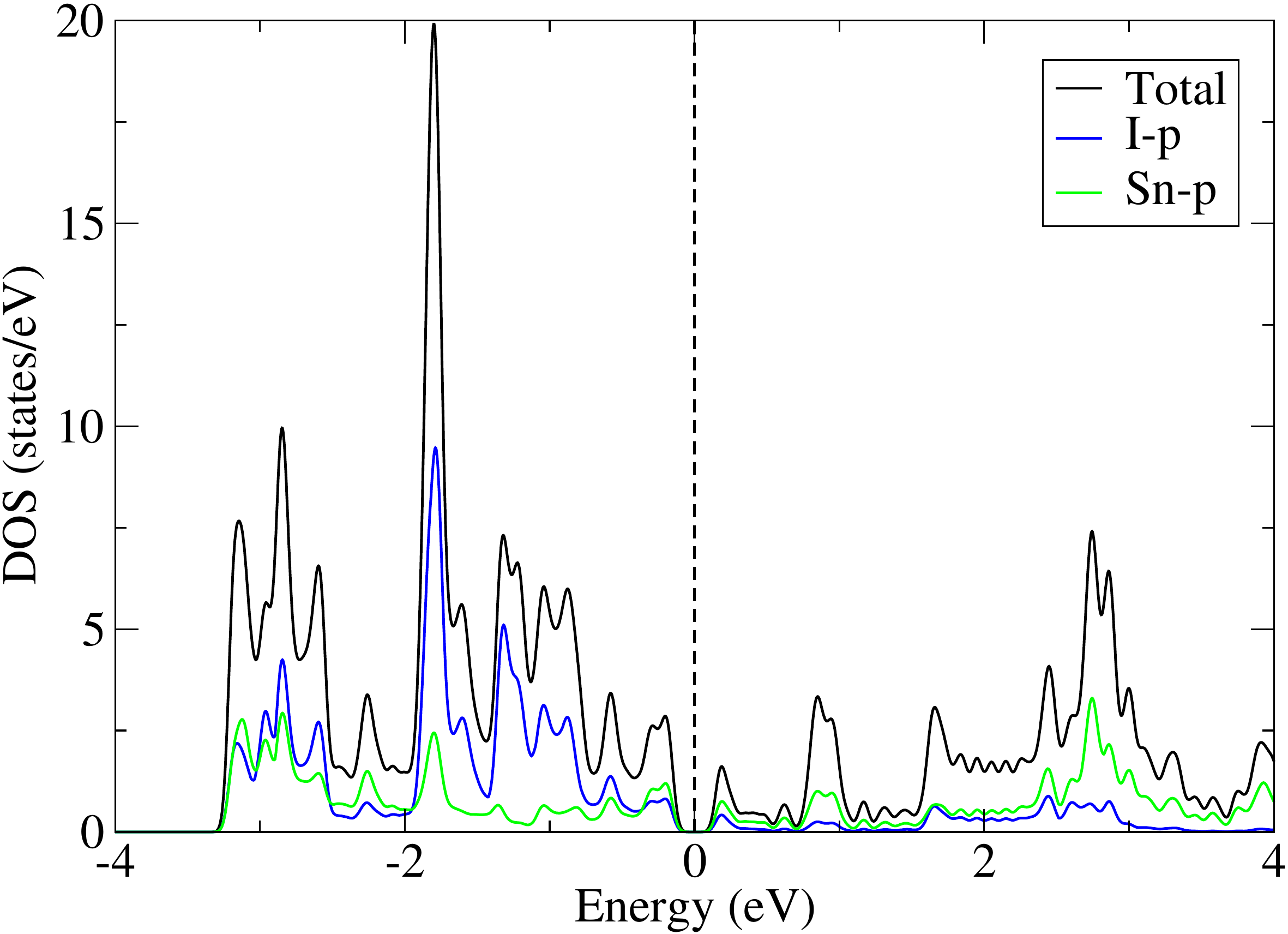}&
\includegraphics[width = 7.5cm, scale=1, clip = true, keepaspectratio]{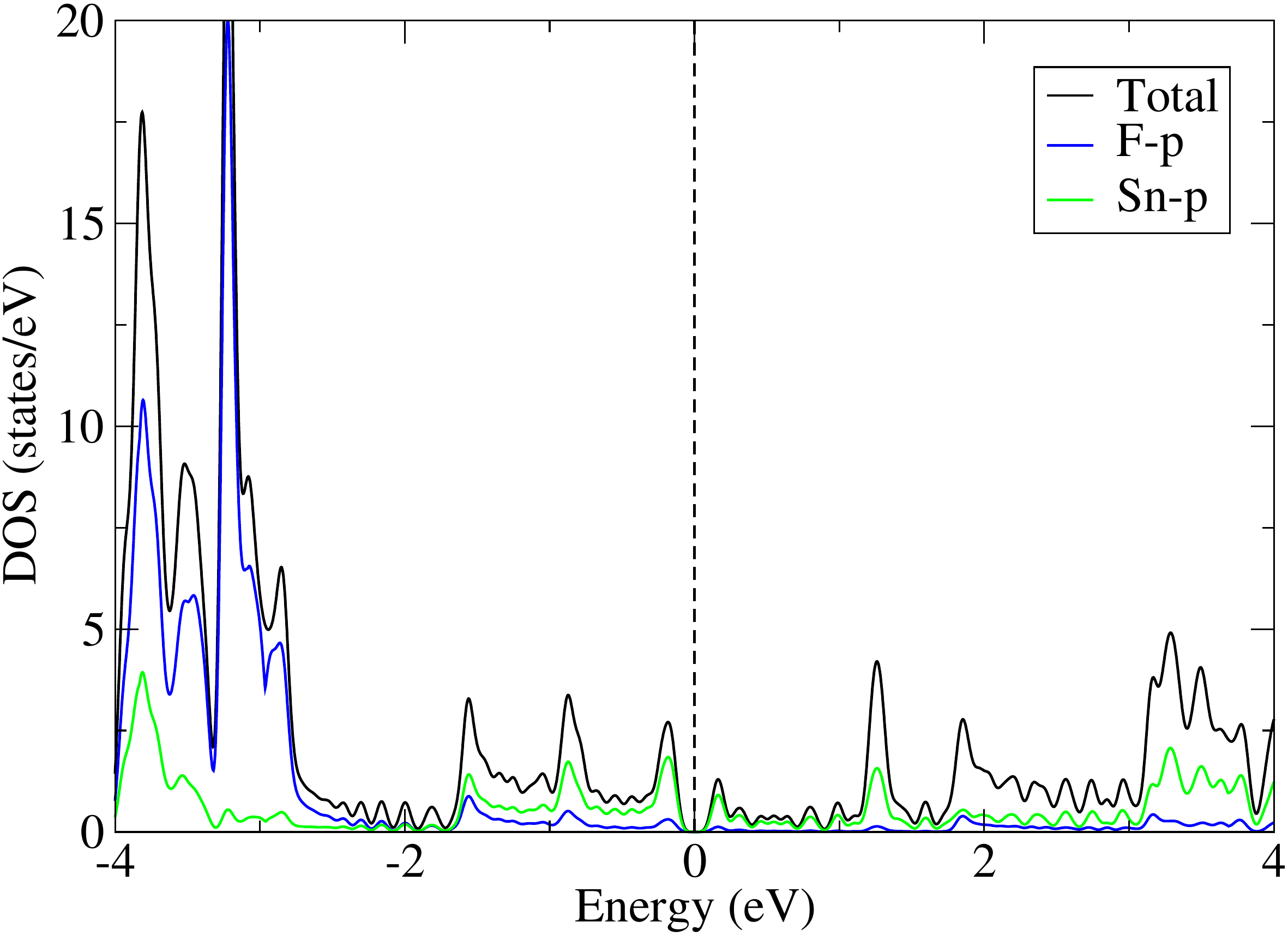}
\end{tabular}
\end{center}
\caption{Atom projected density of states DOS of modified stanene structures. a) -H, b) -OH, c) -CH$_3$, d) -C$_2$H, e) -I and f) F. The zero of energy is set at zero.}
\label{fig:projected_dos}
\end{figure}

In order to
understand the interaction between ligand and the substrate, we have
calculated the band structure, as shown in
Fig.\,\ref{fig:projected_bands}(a)-(f).  The projected DOS is shown in
Fig.\ref{fig:projected_dos}. Bare stanene have a metallic character and zero gap
with a linear dispersion at the $\Gamma$ point, as it is seen in
Fig.\ref{fig:projected_bands} (a). All organic groups can open a gap in stanene. Our results are in
agreement with the ones reported in
Refs.\,\cite{Zhang_2015,Zhang2016,tang2014stable}. Energy gaps are reported in
Table\,\ref{table:properties}.

The band structure for Sn-H is shown in
Fig.\,\ref{fig:projected_bands}(a) for Sn-H,
Fig.\,\ref{fig:projected_bands}(b) for -OH,
Fig.\,\ref{fig:projected_bands}(c) for ${\rm Sn-CH_3}$,
Fig.\,\ref{fig:projected_bands}(d), Sn-C$_2$H,
Fig.\,\ref{fig:projected_bands}(e) for -I and
Fig.\,\ref{fig:projected_bands}(f) for -F.

Contributions to the states close to the Dirac point are mainly due to
Sn-$p$ orbitals. Upon ligand adsorption on stanene, Rashba spliiting
is more emphasized for -OH, I and F groups. I and F are more
electronegative  and therefore induces band gap opening. This
means that the ligand character is also important to determine the
bond strength and consequently the electronic structure of the hybrid
system.  We can clearly see contributions from the functional groups
at VBM and CBM, implying that we have formation of bonds between the
ligand and tin. 

\begin{figure*}[ht!]
\centering
\includegraphics[width=8.0cm,clip]{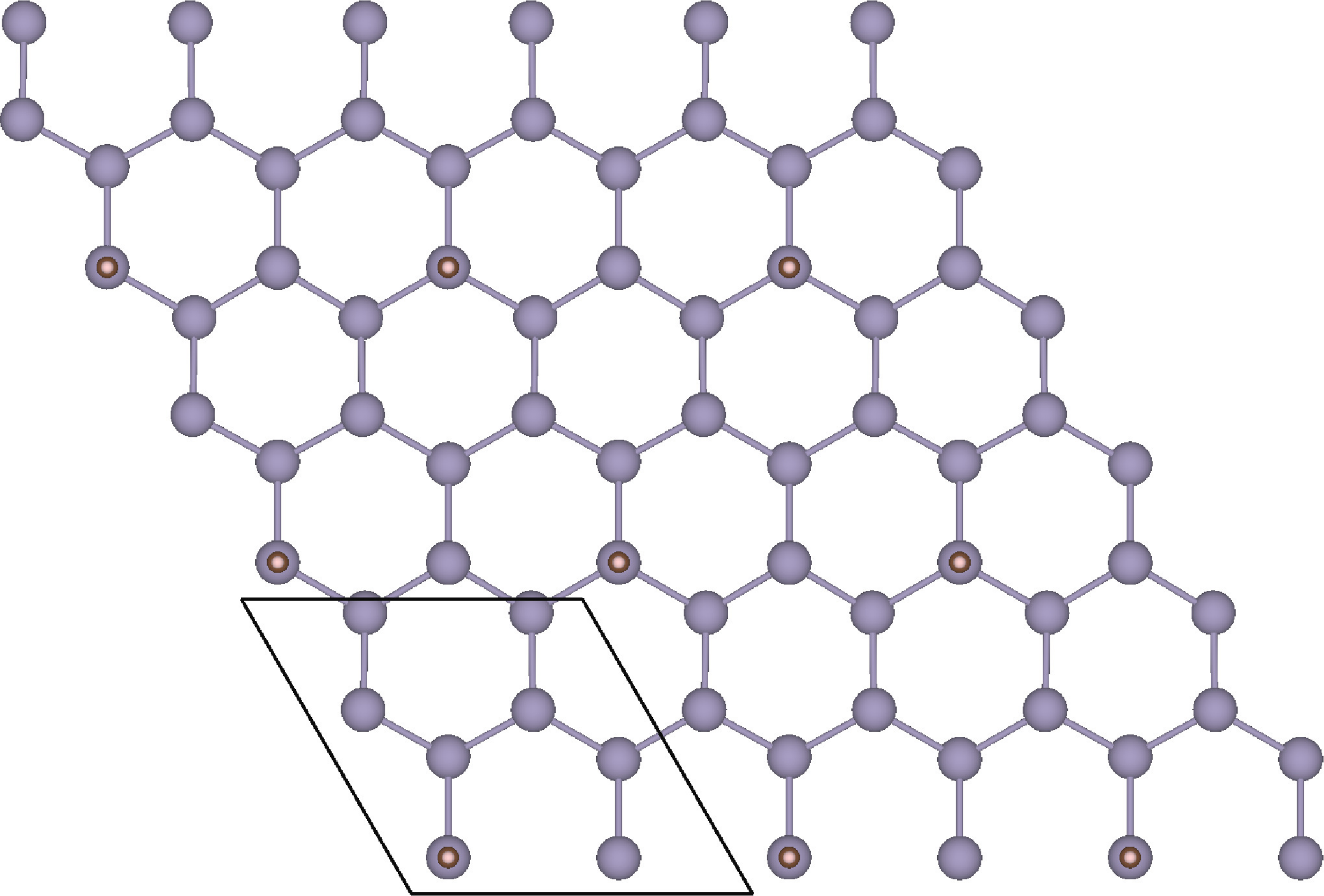}
\includegraphics[width=7.0cm,clip]{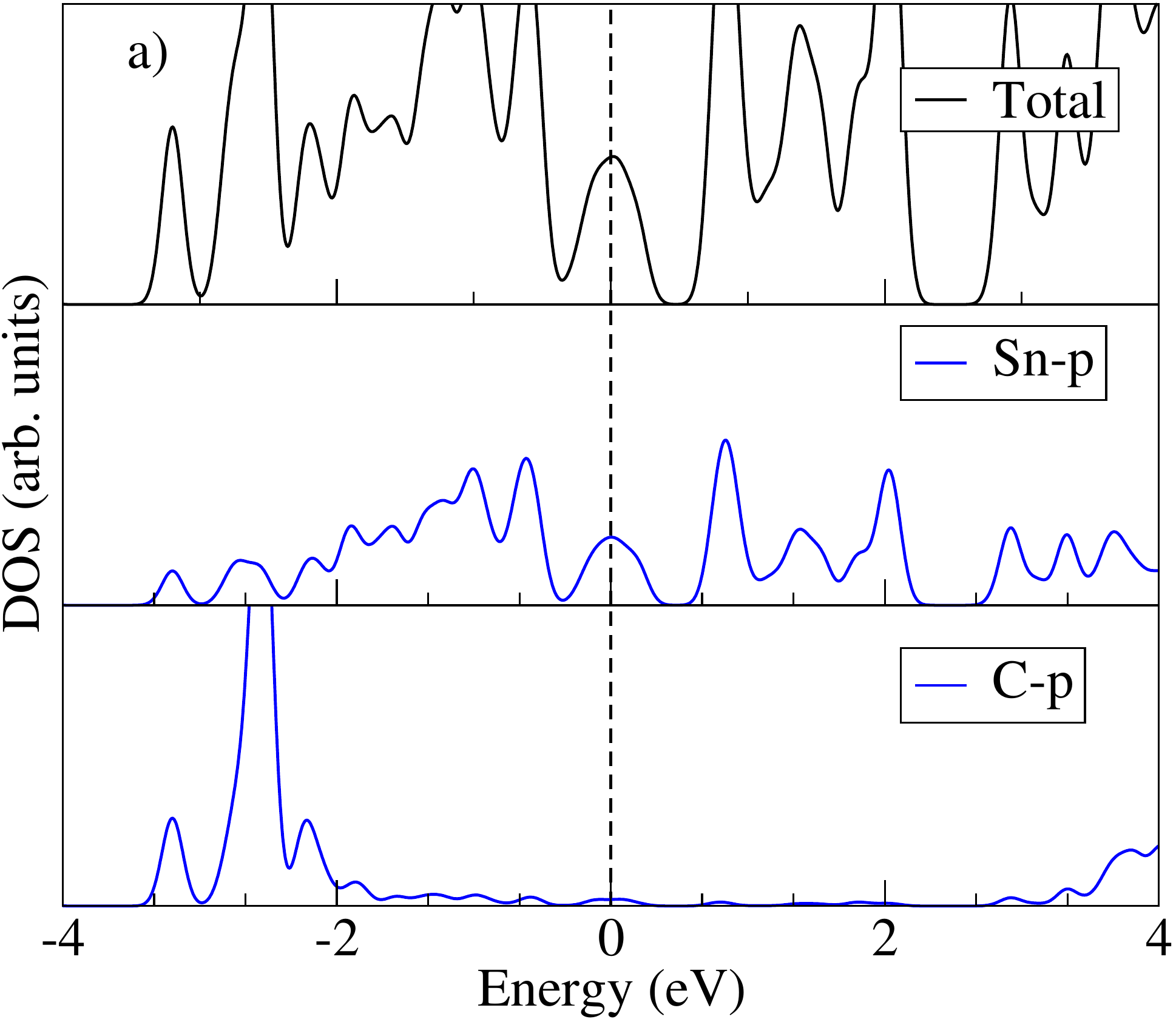}\\
\includegraphics[width=8.0cm,clip]{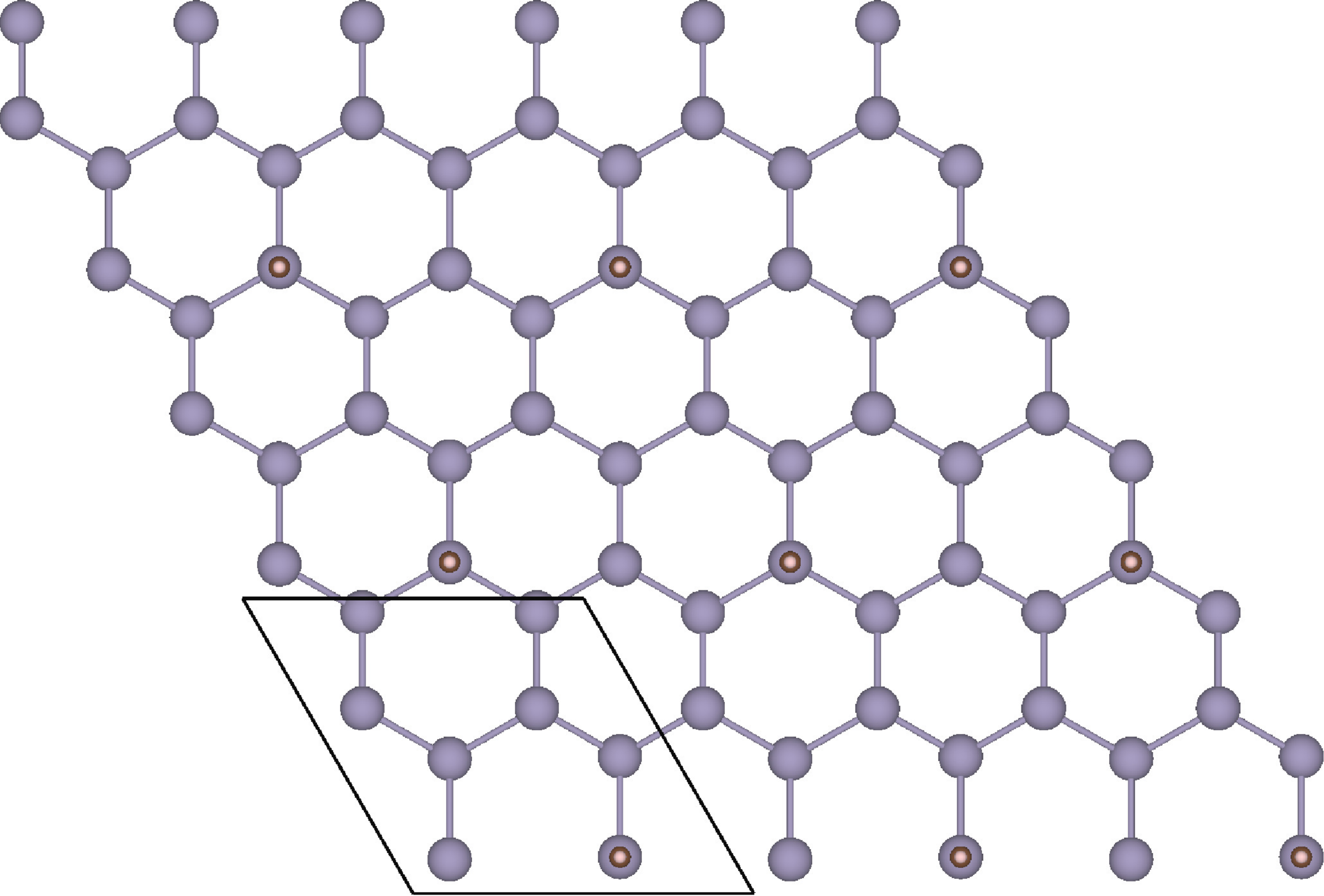}
\includegraphics[width=7.0cm,clip]{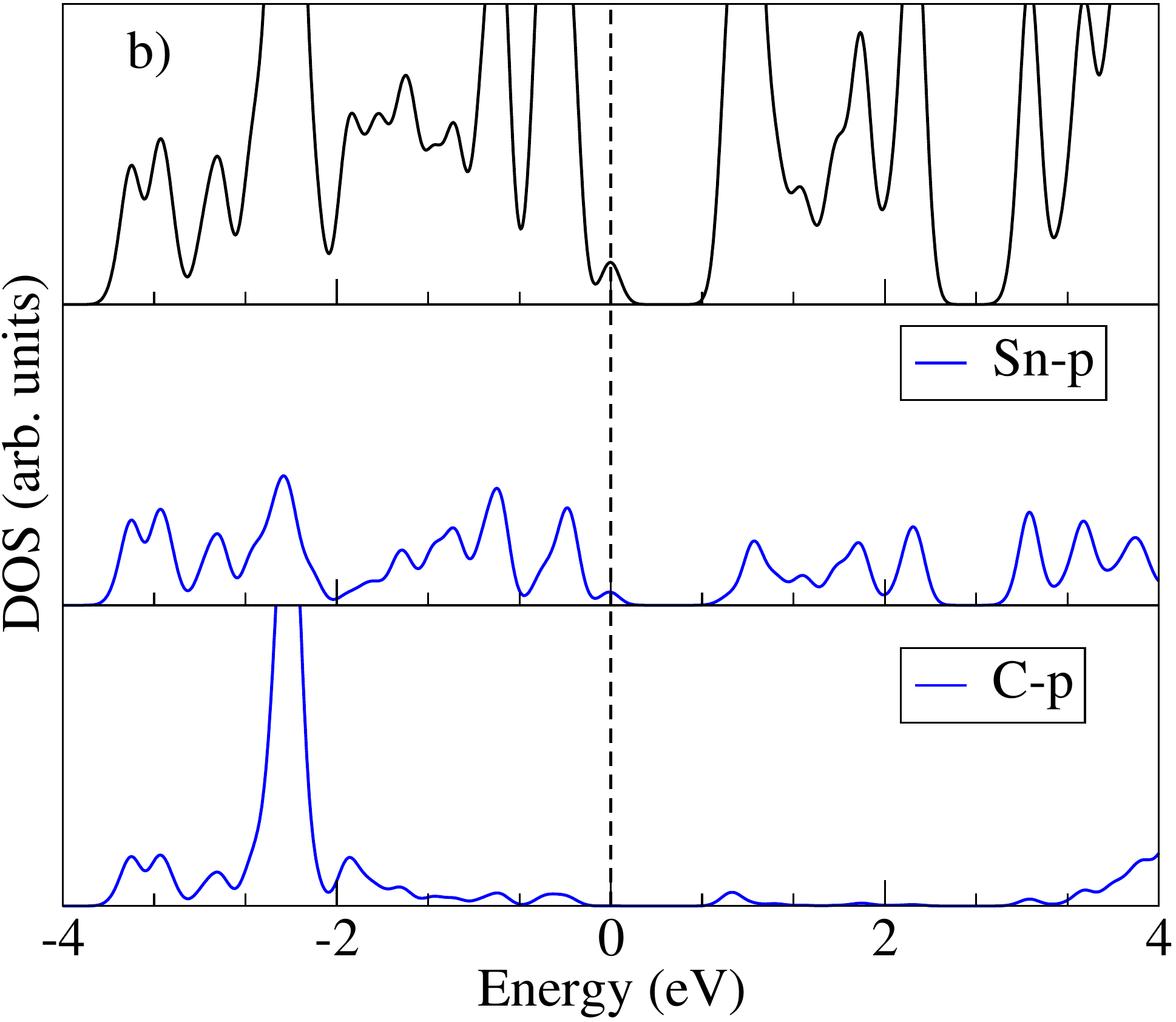}\\
\includegraphics[width=8.0cm,clip]{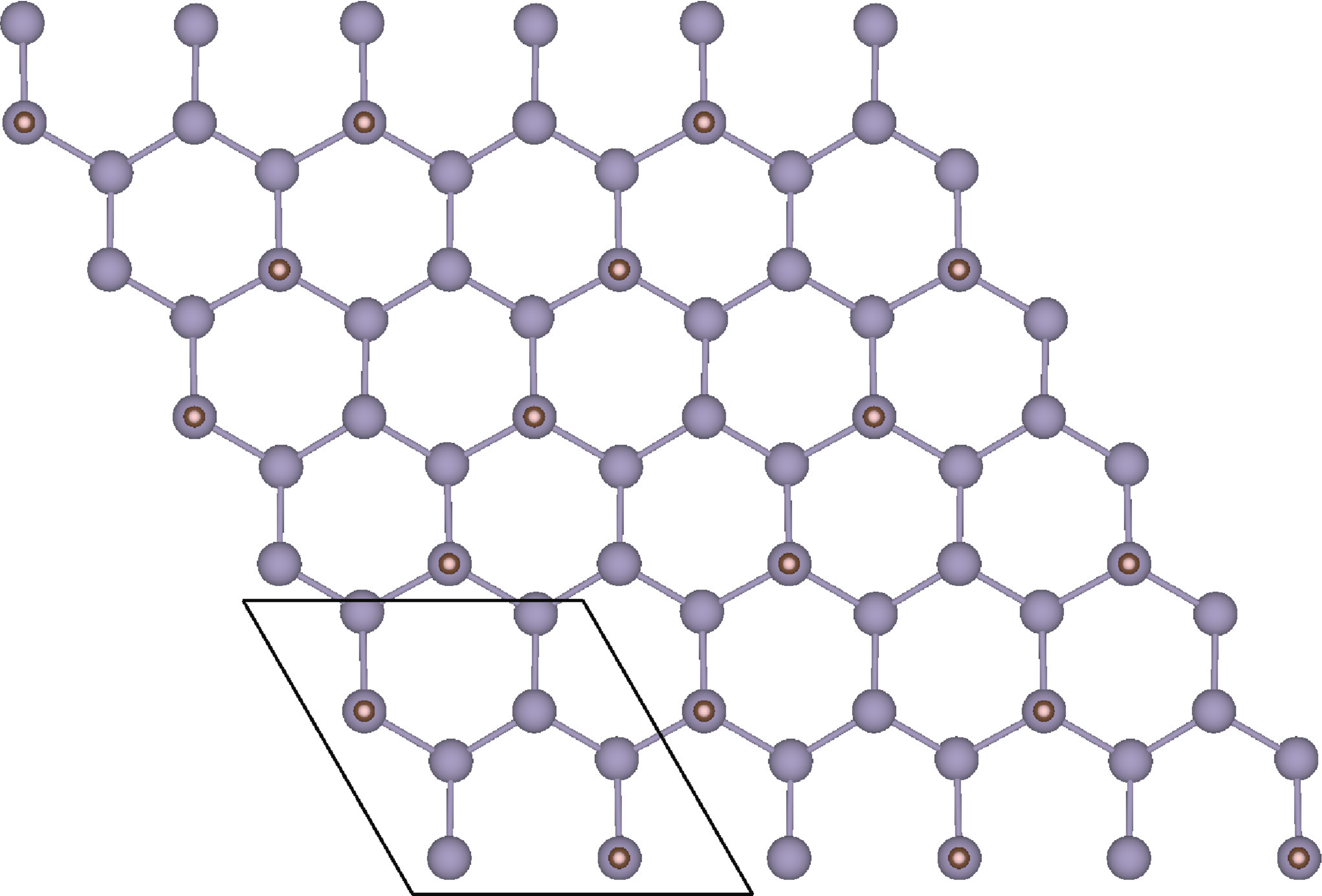}
\includegraphics[width=7.0cm,clip]{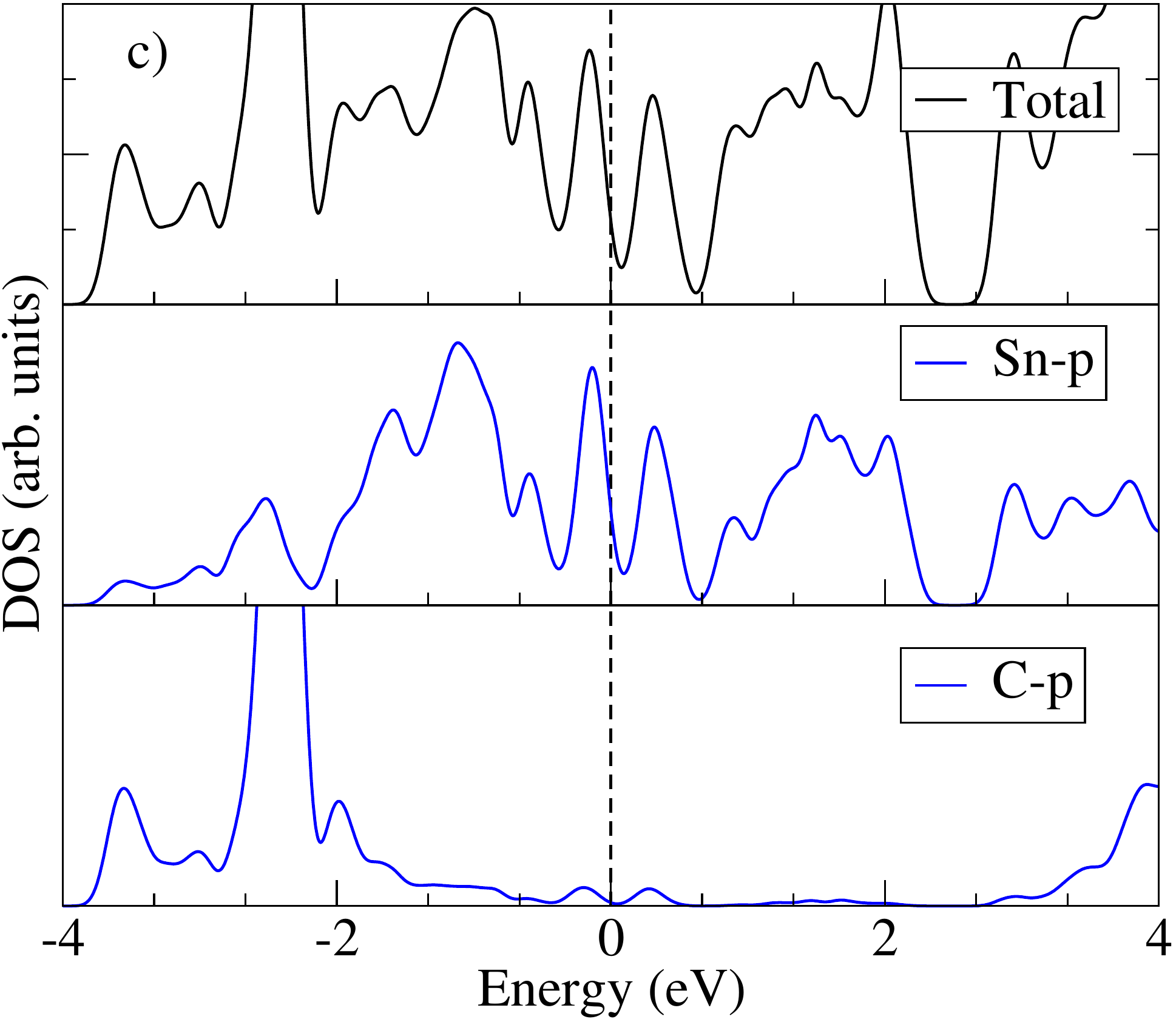}
\caption{\label{fig:partialcoverage} Top views Orbital projected DOS of hybrid Sn-C$_2$H layers
    at: a) 1/8 ML regime with one molecule, b) 1/4 ML regime with two molecules on opposite sides, c) 1/4 ML regime with two molecules on the same side. Brown, blue and white spheres are carbon, hydrogen and tin atoms, respectively. Fermi level is set at zero.}
\end{figure*}

In order to verify the influence of ligand coverage on the stanene
band structure, we have further investigated adsorption at different
ligand-ligand separations. We employ a $(2\times2)$ unit cell with
three different geometries. As a representative system we consider the
Sn-C$_2$H structure. In Fig.\,\ref{fig:partialcoverage} we show the
top views of hybrid Sn-C$_2$H layers with partial coverage. In
Fig.\,\ref{fig:partialcoverage} (a) we show a $(2\times2)$ supercell
at 1/8 ML coverage regime with a single molecule, in
Fig.\,\ref{fig:partialcoverage} (b) at 1/4 ML regime with two
molecules on the opposite sides of stanene,
Fig.\,\ref{fig:partialcoverage} (c) at 1/4 ML coverage regime with two
molecules on the same side. All these structures should be
representative of possible realisable experiments.  As a matter of
comparison, structures with molecules adsorbed on one side should
represent functionalization of stanene grown on a
substrate for example\,\cite{ni2017germanene} whereas structures with molecules
adsorbed on both sides can be achieved by deintercalation, a method
proposed for germanium layers\,\cite{JiangNT:2014}.

The corresponding electronic projected DOS are shown in
Figs.\,\ref{fig:partialcoverage}(a)-(c). All systems have a metallic
behavior. This means that not all configurations and coverages would
lead to band gap opening. It would be necessary to have ligands more
densely packed to open a small, but still sizeable band gap in
stanene. We can therefore conclude that the ligand-ligand interaction
does play a role in the band gap opening of stanene.

The dielectric function of hybrid stanene-organics layers were
investigated by calculating the imaginary part of the dielectric
function at GW level. The imaginary part of the dielectric function is
calculated directly from the electronic structure through the joint
density of states and the momentum matrix elements occupied and
unoccupied eigenstates according Ref.\,\cite{Shishkin:07}.

\begin{figure}[htbp!]
\begin{tabular}{cc}  
\includegraphics[width = 7cm,scale=1, clip = true, keepaspectratio]{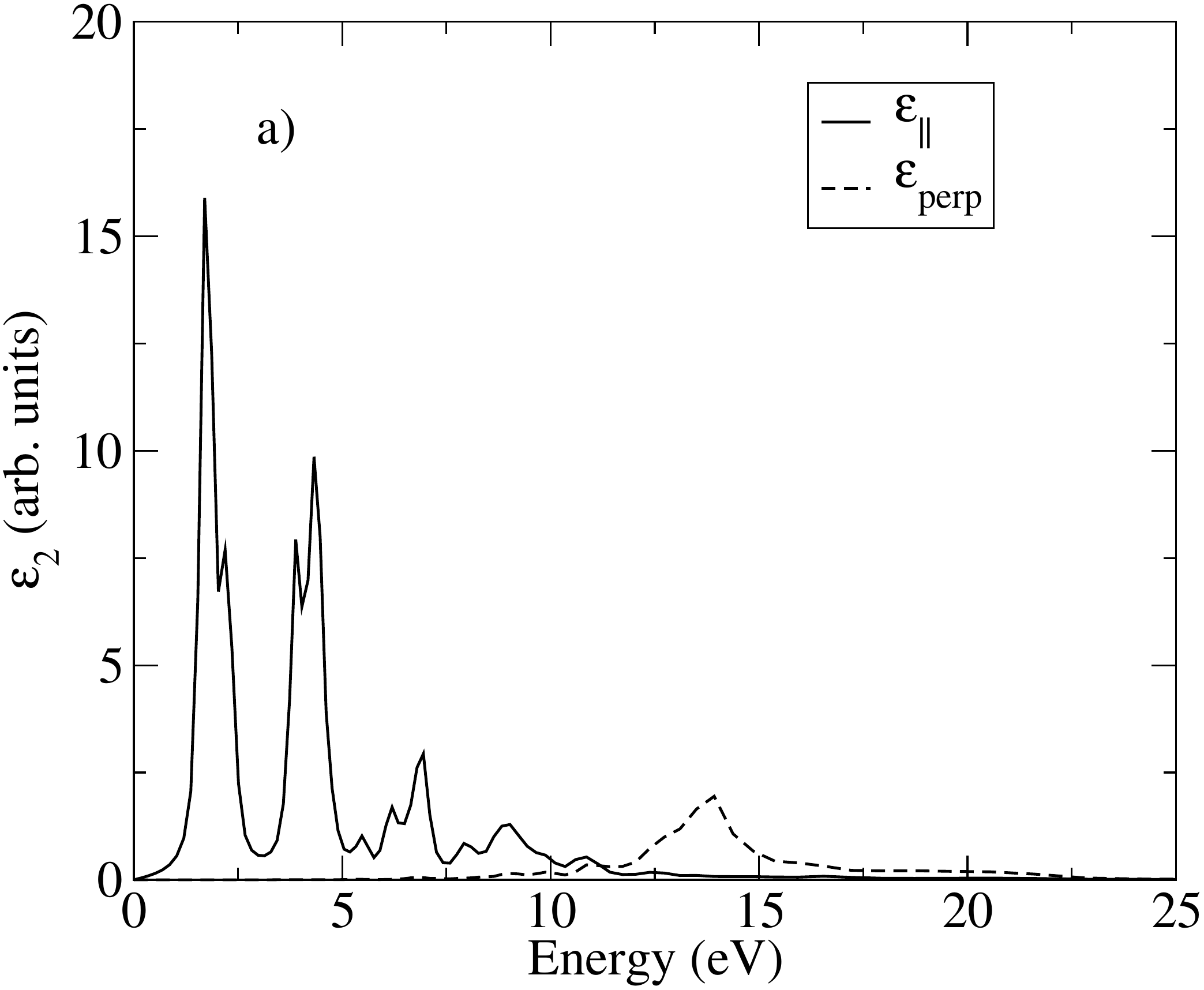}&
\includegraphics[width = 7cm,scale=1, clip = true, keepaspectratio]{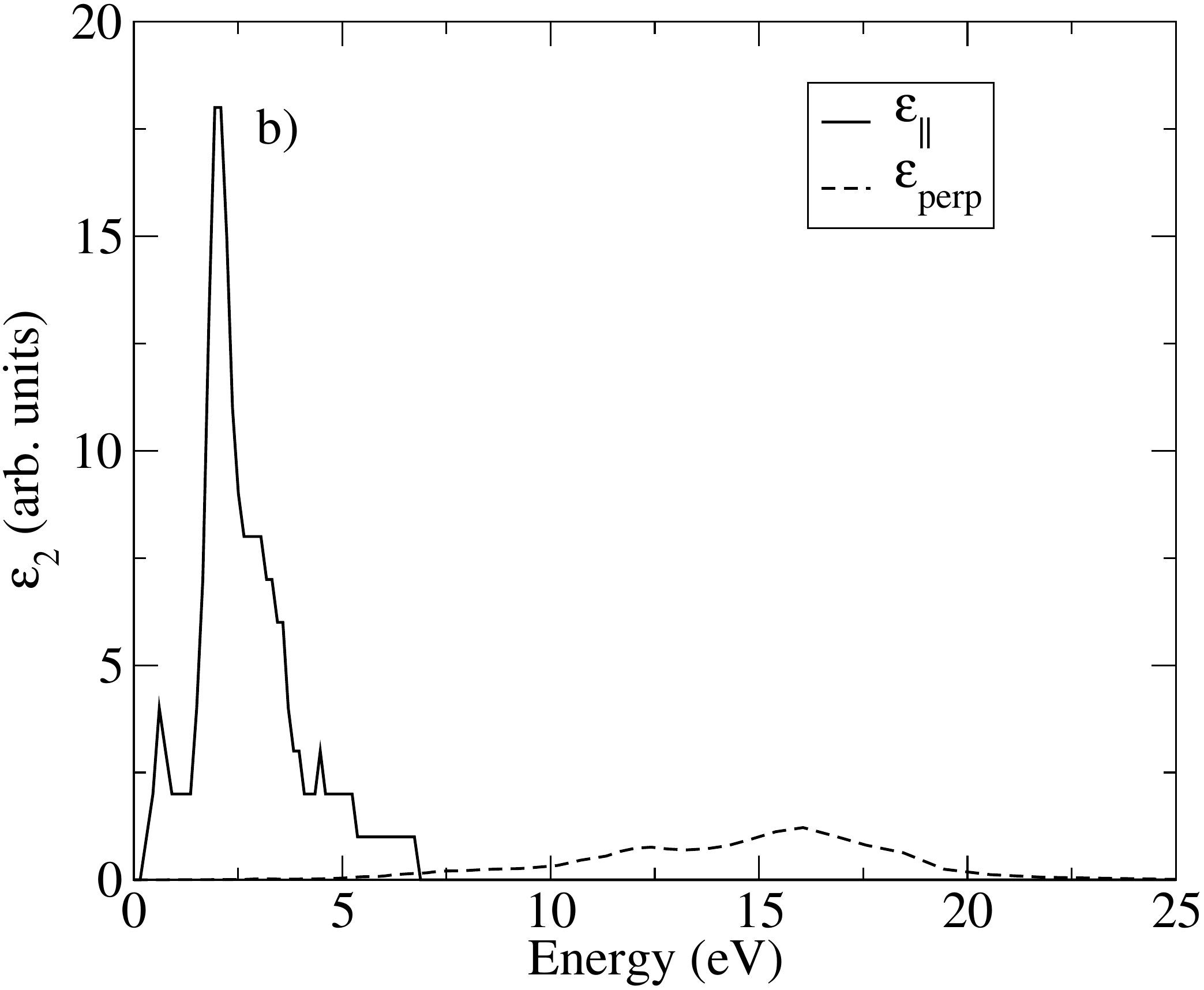}\\
\includegraphics[width = 7cm,scale=1, clip = true, keepaspectratio]{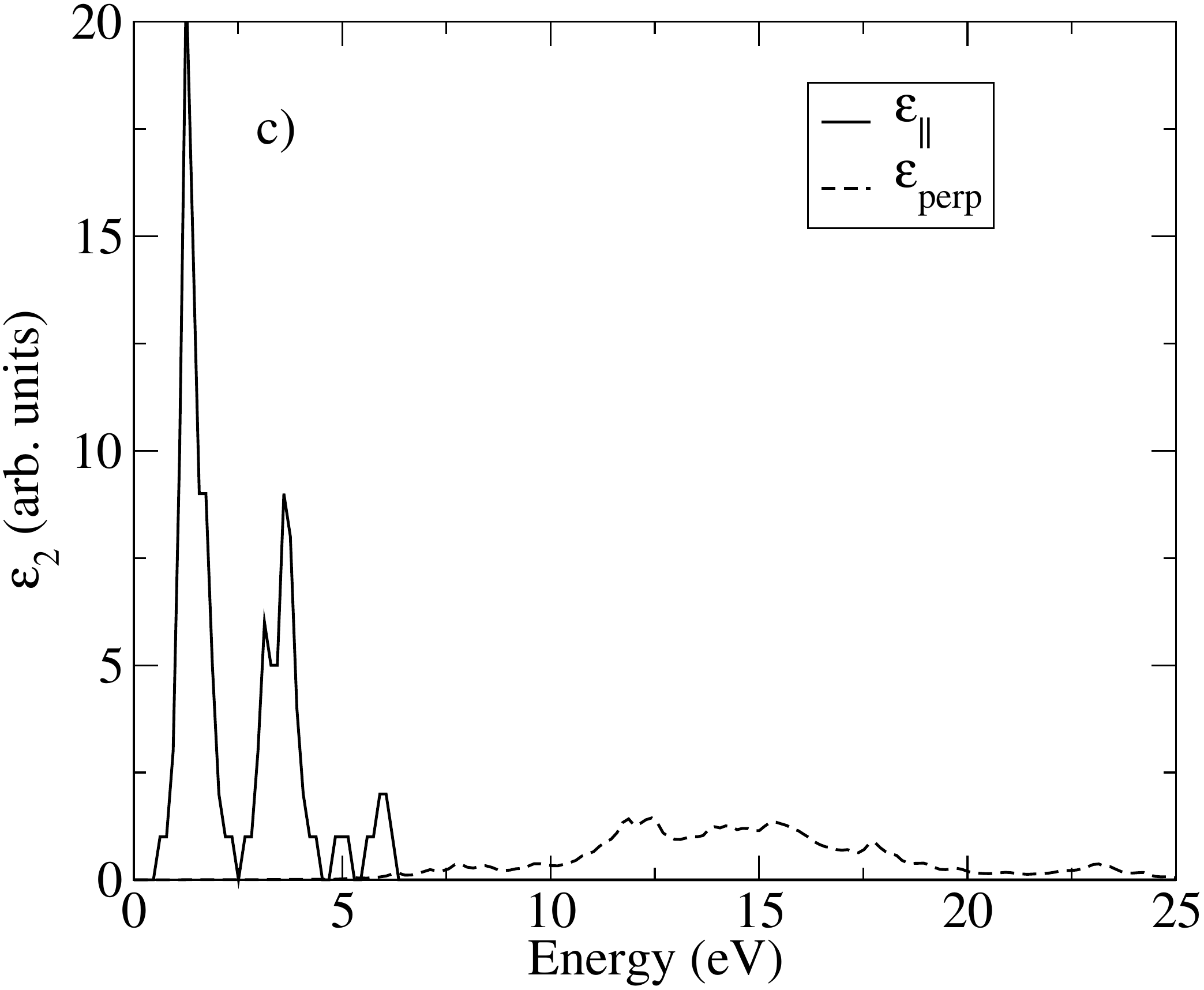}&
\includegraphics[width = 7cm,scale=1, clip = true, keepaspectratio]{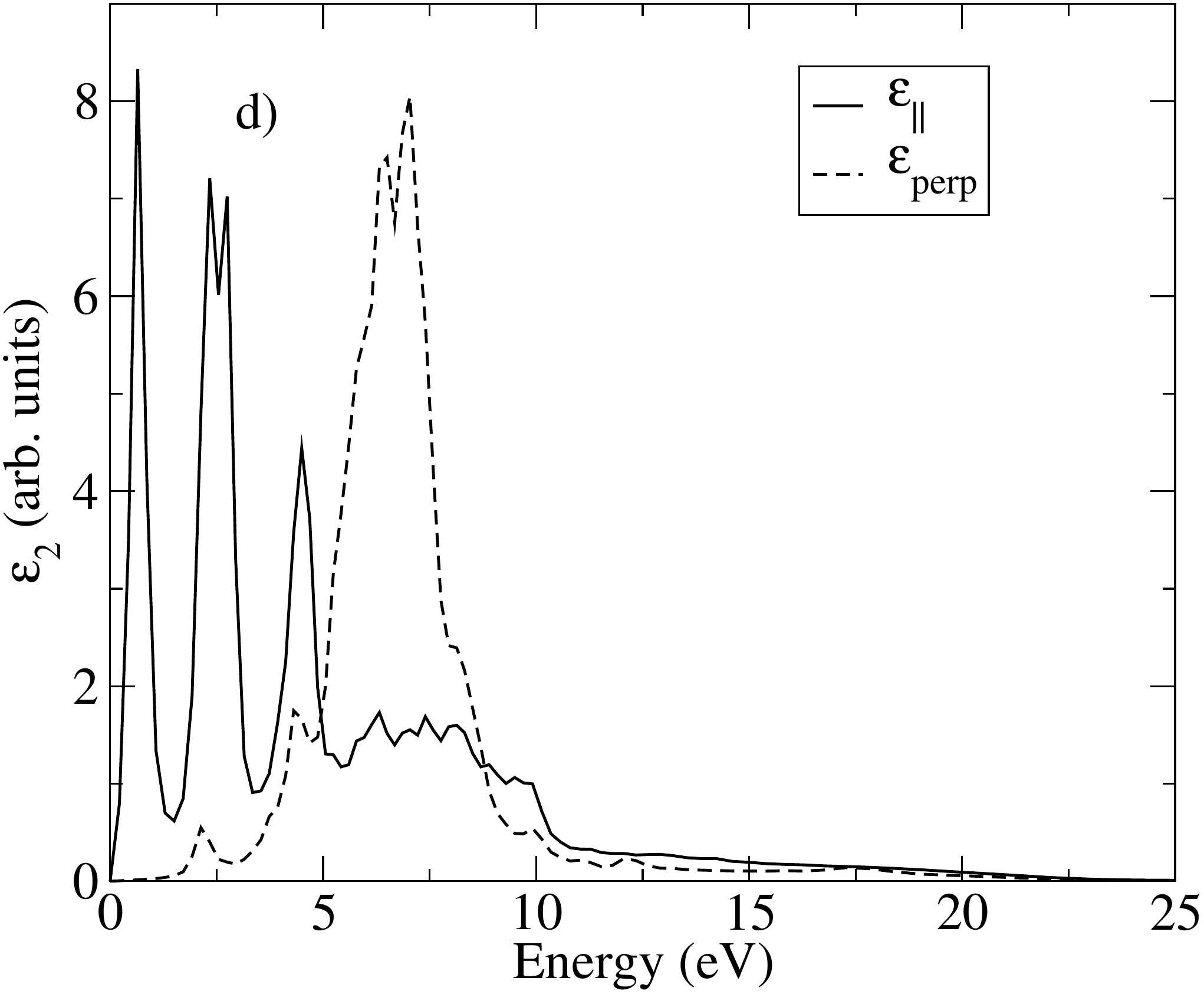}\\
\includegraphics[width = 7cm,scale=1, clip = true, keepaspectratio]{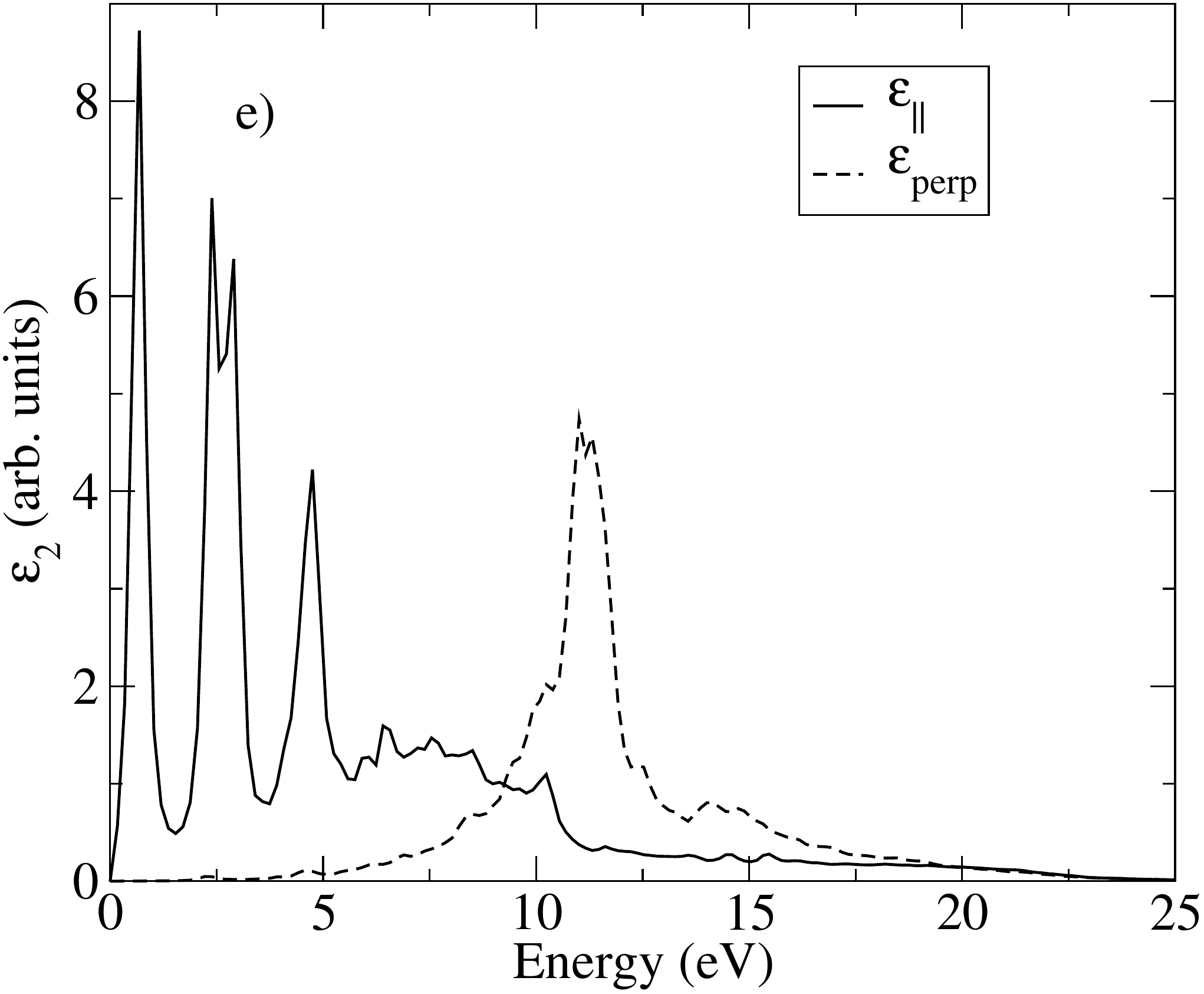}&
\includegraphics[width = 7cm,scale=1, clip = true, keepaspectratio]{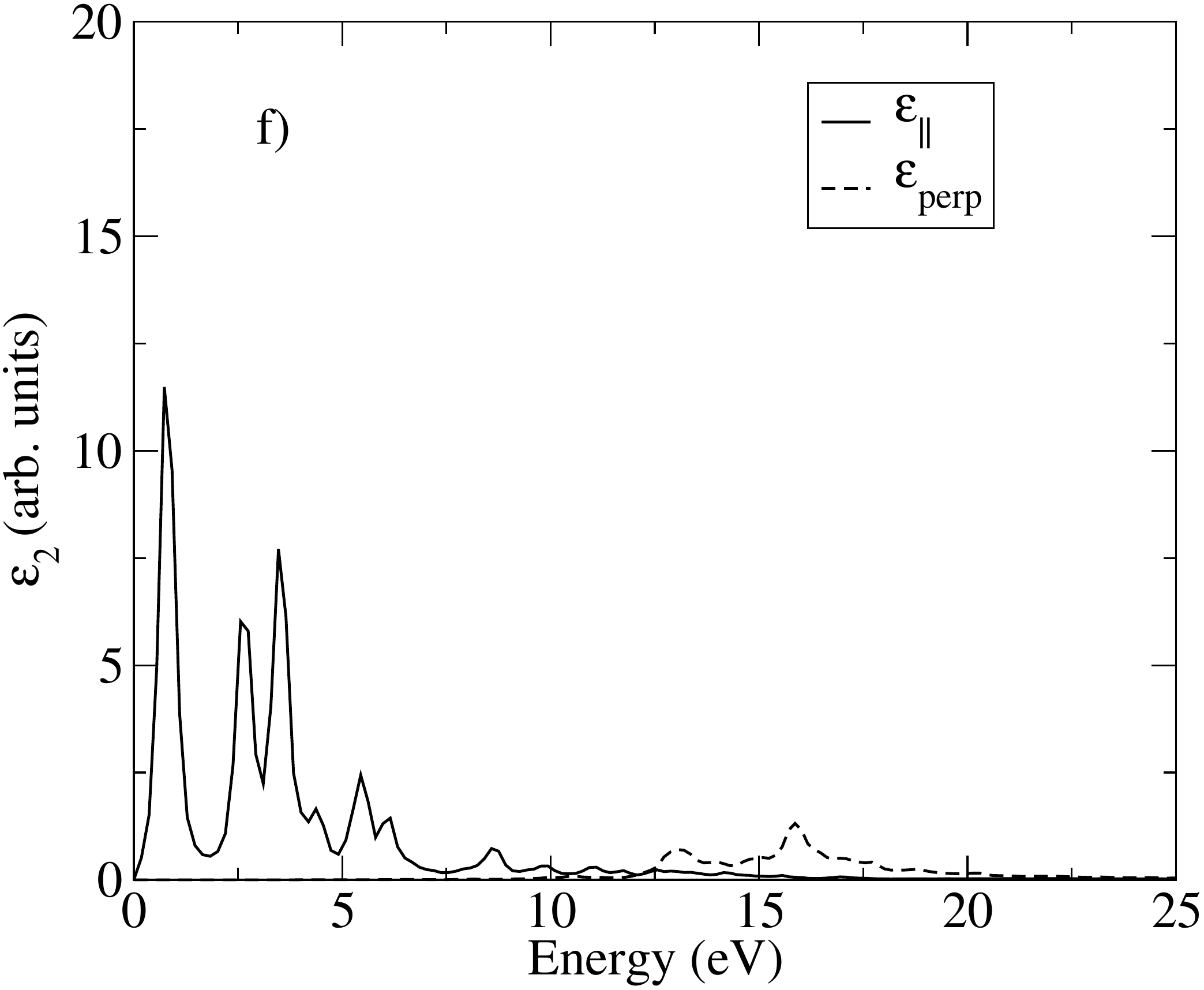}
\end{tabular}
\caption{\label{fig:diel_GW} Imaginary part of the dielectric function for pure and 
  functionalized stanene within GW$_0$ approximations. a) H, b) -OH, c) -CH$_3$, d) -C$_2$H, e) -I and f) -F. Light propagation parallel and perpendicular to the tin layer   is denoted as $\epsilon_{||}=(\epsilon_{xx}+\epsilon_{yy})/2$ and $\epsilon_{\rm perp}=\epsilon_{zz}$, respectively.}
\end{figure}

The bang gaps obtained within GGA, HSE and GW$_0$@HSE are reported in
Table \ref{table:GW}.  One noticed that there is an improvement for H, CH3, C2H
for HSE compared to GGA.  GW compared to HSE there is an improvement
for H, CH3, C2H and a bit for I and F and none for OH.

We show the dielectric function calculated within the G$_0$W$_0$ and
GW approximations in Fig.\,\ref{fig:diel_GW}. The averaged parallel
$\varepsilon_{\|} = (\varepsilon_{\rm xx}+\varepsilon_{\rm yy})/2$ and
perpendicular $\varepsilon_{\perp} = \varepsilon_{zz}$ components of
the the imaginary part of the dielectric function $\varepsilon_2$ are
shown. The $\varepsilon_{\|}$ component corresponds to the propagation
of the external electromagnetic field parallel to the stanene plane
while $\varepsilon_{\perp}$ corresponds to the field perpendicular to
the plane. Because of optical selection rules, anisotropy in the
optical spectra is seen. Anisotropy has also been reported in layered
monochalcogenide of tin sulfide (GeS) \cite{GeS}, black
phosphorous\,\cite{blackP} and bismuthene\,\cite{Ciraci2019,JPCC2020}.
The systems with a gap show finite absorption limits for both parallel
and perpendicular directions with larger intensity for the
$(\varepsilon_{\|}$ component.

As discussed in Ref.\cite{JiangNT:2014}, both ligand size and
electronegativity can change the bond length and band gap of
functionalized germanene.  Larger ligands are expected to lead to
larger Sn-Sn separation, thus yielding a lower band gap. Ligands with
greater electronegativity are expected to withdraw electrons and
therefore lower the band gap. The general conclusion is that ligands
that are more electron-withdrawing and have greater steric bulk will
expand the Sn-Sn framework and lower the band gap, with complete
ligand coverage.

The bare stanene shown in Fig. \ref{fig:diel_GW}(a) has metallic
behavior.

Except for -OH, I and -F groups, HSE does not lead to improvement in
the band gap.  Interestingly enough, water does not dissociate on
stanene\,\cite{chen2016}. The water adsorption is stable on stanene,
as reported in Ref.\cite{Yun2017}, which could explain the low
formation energy of -OH groups. On the other hand, -OH groups are
still able to modify the band structure of stanene by opening a large
gap of 0.22 eV. Besides, -OH groups were found to lead to Quantum Spin
Hall (QSH) effects in stanene\,\cite{Binghai2013}, making them
promising for optoelectronic devices. GW calculations were found to
lead to further improvement compare to HSE for H, CH$_3$, C$_2$H, I and F.
Larger improvement for covalent type bonds such as for H, CH$_3$ and C$_2$H.

As a general feature, two main absorption peaks appear in the spectrum
for all groups, except for -OH. Sn-H shown in
Fig.\,\ref{fig:diel_GW}(a) has peaks at 1.64 and 4.21\,eV. The Sn-OH
spectrum shown in Fig.\,\ref{fig:diel_GW}(b) has peak at 2.05 eV.  A
first peak at 1.42 for Sn-CH$_3$ is seen in
Fig.\,\ref{fig:diel_GW}(c). A second peak appears at 3.48 eV. For
-C$_2$H one has peaks at 0.75 and 2.50 eV as it can be seen in
Fig.\,\ref{fig:diel_GW}(d). -I and -F have peaks at similar positions,
as seen in Figs.\,\ref{fig:diel_GW}(e) and (f). This could be due the
similar electronegativity

\begin{table*}[ht!]
\begin{tabular*}{1.0\textwidth}{@{\extracolsep{\fill}}lcccc}
\hline
ligand  & \multicolumn{2}{c}{E$_{gap}$(eV)} & 1st peak & 2nd peak \\
       &          HSE   &  GW$_0$     &               &   \\
\hline
-H        &  0.50 &  1.10  & 1.64 & 4.21  \\ 
-OH       &  0.19 &  0.18  & 2.05 & -  \\
-CH$_3$   &  0.37 & 1.05  & 1.42 & 3.48  \\ 
-C$_2$H   &  0.38 & 0.62  & 0.75 & 2.50   \\ 
-I        &  0.34 & 0.60  & 0.70 & 2.60     \\ 
-F        &  0.30 & 0.62  & 0.75 & 2.61   \\
\hline
\end{tabular*}
\caption{Gw$_0$ band gap $E_{gap}$ and first and second energy transitions of functionalized stanene.}
\label{table:GW}
\end{table*}

\section{Conclusions}

We have performed first-principles calculations of stanene
functionalized layers with small organic groups. Our charge density
analysis show that the ligands are chemisorbed on the tin layers. Our
calculations for the dielectric properties of bare and ligand adsorbed
stanene show a large anisotropy and that the absorption onset is
determined by ligand electronegativity. We
believe our findings of a finite gap shows open a path for rational
design of nanostructures with possible applications in biosensors and
solar cells.

\section{Acknowledgements}

We acknowledge the financial support from the Brazilian Agency CNPq
and German Science Foundation (DFG) under the program FOR1616. The
calculations have been performed using the computational facilities of
Supercomputer Santos Dumont and at QM3 cluster at the Bremen Center
for Computational Materials Science and CENAPAD.

\bibliographystyle{apsrev}

\end{document}